\documentclass{aa}
\usepackage{psfig}
\usepackage{epsfig}

\newcommand{\cmsq}{{\rm atoms\,cm}^{-2}}
\newcommand{\ergs}{{\rm erg\,s}^{-1}}

\newcommand{\ergcms}{{\rm erg\,cm}^{-2}{\rm s}^{-1}}
\newcommand{\ltap}{\mathrel{\hbox{\rlap{\lower.55ex \hbox {$\sim$}}
                   \kern-.3em \raise.4ex \hbox{$<$}}}}
\newcommand{\gtap}{\mathrel{\hbox{\rlap{\lower.55ex \hbox {$\sim$}}
                   \kern-.3em \raise.4ex \hbox{$>$}}}}

\begin{document}

\title{Six years of BeppoSAX Wide Field Cameras observations of nine 
galactic type\,I X-ray bursters}

\author{R.\ Cornelisse\inst{1,2} \and  J.J.M.\ in 't Zand\inst{1} 
        \and F.\ Verbunt\inst{2} \and E.\ Kuulkers\inst{3}
        \and J.\ Heise\inst{1} \and P.R. den Hartog\inst{2,1}
        \and M. Cocchi\inst{4} \and L. Natalucci\inst{4}
        \and A. Bazzano\inst{4} \and P. Ubertini\inst{4}}

        \offprints{R.\ Cornelisse}
\mail{R.Cornelisse@sron.nl}

\institute{SRON National Institute for Space Research, Sorbonnelaan 2,
              3584 CA Utrecht, The Netherlands
         \and  Astronomical Institute,
              P.O.Box 80000, 3508 TA Utrecht, The Netherlands
         \and ESA-ESTEC, SCI-SDG, Keplerlaan 1, 2201 AZ Noordwijk, 
              The Netherlands
         \and Istituto di Astrofisica Spaziale (CNR), Area Ricerca Roma
              Tor Vergata, Via del Fosso del Cavaliere, I-00133, Roma, Italy
                }

\date{\today / Accepted date}

\abstract{We present an overview of BeppoSAX Wide Field Cameras
  observations of the nine most frequent type\,I X-ray bursters in the
  Galactic center region. Six years of observations (from 1996 to
  2002) have amounted to 7 Ms of Galactic center observations and the
  detection of 1823 bursts. The 3 most frequent bursters are
  GX\,354$-$0 (423 bursts), KS\,1731$-$260 (339) and GS\,1826$-$24
  (260). These numbers reflect an unique dataset. We show that all
  sources have the same global burst behavior as a function of
  luminosity.  At the lowest luminosities ($L_{\rm
  X}$$\ltap$$2\times10^{37}$ $\ergs$) bursts occur quasi-periodically
  and the burst rate increases linearly with accretion rate (clear in
  e.g. GS\,1826$-$24 and KS\,1731$-$260). At $L_{\rm
  pers}$=2$\times10^{37}$ $\ergs$ the burst rate drops by a factor of
  five. This corresponds to the transition from, on average, a
  hydrogen-rich to a pure helium environment in which the flashes
  originate that are responsible for the bursts. At higher
  luminosities the bursts recur irregularly; no bursts are observed at
  the highest luminosities.  Our central finding is that most of the
  trends in bursting behavior are driven by the onset of stable
  hydrogen burning in the neutron star atmosphere. Furthermore, we
  notice three new observational fact which are difficult to explain
  with current burst theory: the presence of short pure-helium bursts
  at the lowest accretion regimes, the bimodal distribution of peak
  burst rates, and an accretion rate that is ten times higher than
  predicted at which the onset of stable hydrogen burning occurs.
  Finally, we note that our investigation is the first to signal
  quasi-periodic burst recurrence in KS\,1731-260, and a clear
  proportionality between the frequency of the quasi-periodicity and
  the persistent flux in GS\,1826-24 and KS\,1731-260.
  }

\titlerunning{six years of Wide Field Cameras observations} 

\maketitle

\section{Introduction}
Since the discovery of type\,I bursts by Grindlay \& Heise (1975)
about 65 other X-ray bursters have been discovered (e.g., in 't Zand
2001). Most of these are concentrated toward the Galactic center,
which illustrates their Galactic origin. X-ray bursts are characterized by
a fast rise and an exponential decay with durations ranging from
seconds to tens of minutes. Their spectrum can best be described by
black body radiation with cooling during the decay of the burst. These
type\,I X-ray bursts are due to unstable hydrogen/helium burning in a
thin shell on a neutron star surface (see, e.g., the review by Lewin
et~al. 1993).

By far most of the X-ray bursts are emitted by sources, persistent or
transient, with luminosities of $10^{36-37}$ $\ergs$. Some sources at
higher persistent luminosities also show X-ray bursts (e.g., Kuulkers
et~al. 2002), but such bursts are less common. At lower luminosities
bursts have also been sporadically observed (e.g., Gotthelf \&
Kulkarni 1997, Cocchi et~al. 2001a, Cornelisse et~al. 2002). 

Assuming that the amount of fuel burnt per burst is roughly the same,
one expects that the burst rate increases linearly with accretion
rate. However, for most X-ray bursters where it is possible to study
this the opposite is observed (van Paradijs et~al. 1988a), see e.g.
GX\,3$+$1 (den Hartog et~al. 2002).  Bildsten (2000) noted that the
onset of a burst is governed by the local rather than the global
accretion rate (see also Marshall 1982). If the area on the neutron
star on which accretion takes place increases rapidly with the global
accretion rate, the local accretion rate may actually drop, giving
rise to a lower burst rate.

Observations of several sources (e.g., 4U\,1705$-$44, Gottwald et~al.
1989; EXO 0748$-$676, Gottwald et~al. 1986), show different burst
properties at different accretions rates.  Fujimoto et~al. (1981)
predicted this behavior by showing that the composition of the unstable
burning shell changes with accretion (see also Bildsten 1998 for a
recent overview).  Briefly, at the highest accretion rates ($\dot M
\gtap 10^{-9}$ M$_\odot$ yr$^{-1}$) the helium ignites in an unstable
fashion in a mixed He/H environment, causing bursts with durations of
minutes. At the intermediate accretion regime ($10^{-9} \gtap \dot
M/($M$_\odot$ yr$^{-1}$) $\gtap 2\times10^{-10}$) the helium ignites
in an unstable fashion  in a hydrogen-poor environment,
causing bursts with durations smaller than 10 s. At the
lowest accretion regime ($\dot M \ltap 2\times10^{-10}$ M$_\odot$
yr$^{-1}$) unstable hydrogen burning triggers a helium flash causing
bursts with durations larger than 10 s.

If the accretion rate is stable over long periods of time,
periodic burst behavior is expected. It should always take the same
amount of time to accrete enough matter to start the unstable burning
again. This quasi-periodic burst behavior is observed in several burst
sources, for example  4U\,1820$-$30 (Haberl et~al. 1987) or 4U\,1705$-$44 
(Langmeier et~al. 1987), but only for limited periods of time. During
other periods the occurrence of bursts appear completely a-periodic.  An
exception is GS\,1826$-$24 whose bursts are always seen to recur quasi
periodically (Ubertini et~al. 1999; Cocchi et~al. 2001b).

In this paper we describe the burst behavior of the nine most
frequent X-ray bursters in the Galactic center region observed with
the Wide Field Cameras. All are known X-ray bursters and most of them
are persistently bright ($L_{\rm X}$$\sim$$10^{36-37}$ $\ergs$). We
compare these bursters with each other and with other bursters. The
observations and the search for type\,I bursts are described in
Sect.\,2. In Sect.\,3 we present the results. We start in Sect.\,3.1
with the general properties of the nine burst sources. In Sect.\,3.2
we discuss the wait time as a function of persistent emission for the
bursters where this is possible. In Sect.\,3.3 we compare the
exponential decay times of the bursts with the theoretical regimes.
Finally in Sect.\,4.1 we start with a summary of the observations and
compare our results with previous studies and in Sect.\,4.2 discuss
some implications for burst theory. We also derive some general
properties of the population of X-ray bursters.

\section{Observations and data analysis}

The BeppoSAX satellite operated from May 1996 until May 2002 (Boella
et~al. 1997). During this period the Wide Field Cameras (Jager et~al.
1997) onboard the satellite observed the Galactic center region each
spring and fall with an average schedule of one day per week. This
adds up to 12 Galactic center campaigns with a total net observation
time of 7 Ms. The Wide Field Cameras (WFC) are two identical coded
mask cameras with a $40^\circ\times40^\circ$ field of view, a $5'$
angular resolution, 2-28 keV bandpass and 20\% spectral
resolution (full width at half maximum at 6 keV). The large field of view
combined with the good angular resolution makes it an excellent
instrument to simultaneously observe a large fraction (50\%) of the
low mass X-ray binary (LMXB) population in our Galaxy when pointed at
the Galactic center.

Each source in the field of view casts a shadow of the mask pattern
on the detector. The detector accumulates the sum of differently
shifted mask shadows. By cross-correlating this detector image with
the mask pattern a sky image is reconstructed (e.g., in 't Zand
1992). This procedure is supplemented with a dedicated iterative
cleaning algorithm (Hammersley et~al. 1992). When no new sources are
detected in the iterative process the background is estimated from the
(supposedly) empty sky image. Lightcurves are constructed in the full
bandpass for each detected source with a time resolution of 5 s, which
is a trade-off between the average duration of a type\,I burst
($\simeq10$ s) and the statistical quality of the data.

\begin{table}[t]
\caption{Overview of the nine most frequent burster sources in the
Galactic center region. They are ordered in decreasing number of
bursts observed per source with the WFC.  For each source a factor
(conv.) in erg/count is derived to convert photon flux to energy flux
in 2-28 keV . We also show the net exposure for each individual source
and the distance as quoted in the literature. For each distance
estimate we assume an error of 30\%. The distances are derived from:
[1] Galloway et~al. 2002, [2] Muno et~al. 2000, [3] in 't Zand
et~al. 1999, [4] Lutovinov et~al. 2001, [5] Muno et~al. 2001, [6]
Gottwald et~al. 1989, [7] Augusteijn et~al. 1998, [8] Kuulkers \& van
der Klis 2000, [9] Heasley et~al. 2000.
\label{top}}
\begin{tabular}{lccccl}
\hline
\hline
Source & \#bursts & exp. time & conv. & d & ref.\\
       & WFC      &  (Ms) & ($10^{-8}$)  & (kpc) &\\
\hline
GX\,354$-$0    & 423 & 7.4 & 1.9 & 5.4 & [1]\\
KS\,1731$-$260 & 339 & 6.7 & 1.5 & 7.0 & [2]\\
GS\,1826$-$24  & 260 & 6.5 & 1.9 & 8.0 & [3]\\
A\,1742$-$294  & 178 & 7.0 & 2.2 & 8.5 & [4]\\
4U\,1702$-$429 & 104 & 8.9 & 1.7 & 6.7 & [5]\\
4U\,1705$-$44  &  66 & 8.7 & 1.8 & 8.9 & a\\
4U\,1636$-$536 &  61 & 4.7 & 1.5 & 5.9 & [7]\\
GX\,3$+$1      &  61 & 6.9 & 1.6 & 4.5 & [8]\\
4U\,1820$-$30  &  49 & 7.1 & 1.6 & 7.6 & [9]\\
\hline
\multicolumn{5}{l}{a: distance estimated from [6]}\\

\end{tabular}
\end{table}

These lightcurves are searched for X-ray bursts in the following
manner. We estimate the average flux and the standard deviation
($\sigma$) for each orbit (about 60 minutes of net exposure time). If
a bin is at least 4$\sigma$ above the average flux we mark this as a
candidate burst. We estimate that we have, on average, only one false
peak triggered as a burst per observation. A triggered bin is visually
approved as a burst if the shape of the lightcurve around the bin can
be described by a fast rise and exponential decay. If needed, this is
done at 1 s time resolution. Thus, we detect all bursts with e-folding
times $\gtap$2 s and peak fluxes in excess of 0.5 Crab. Bursts with
lower peak intensities or shorter e-folding times do exist but this is
only a minor fraction compared to the bursts detected (e.g. Cocchi
et~al. 2001b, van Paradijs et~al. 1988b).

\begin{figure*}
\parbox[b]{6.0cm}{\psfig{figure=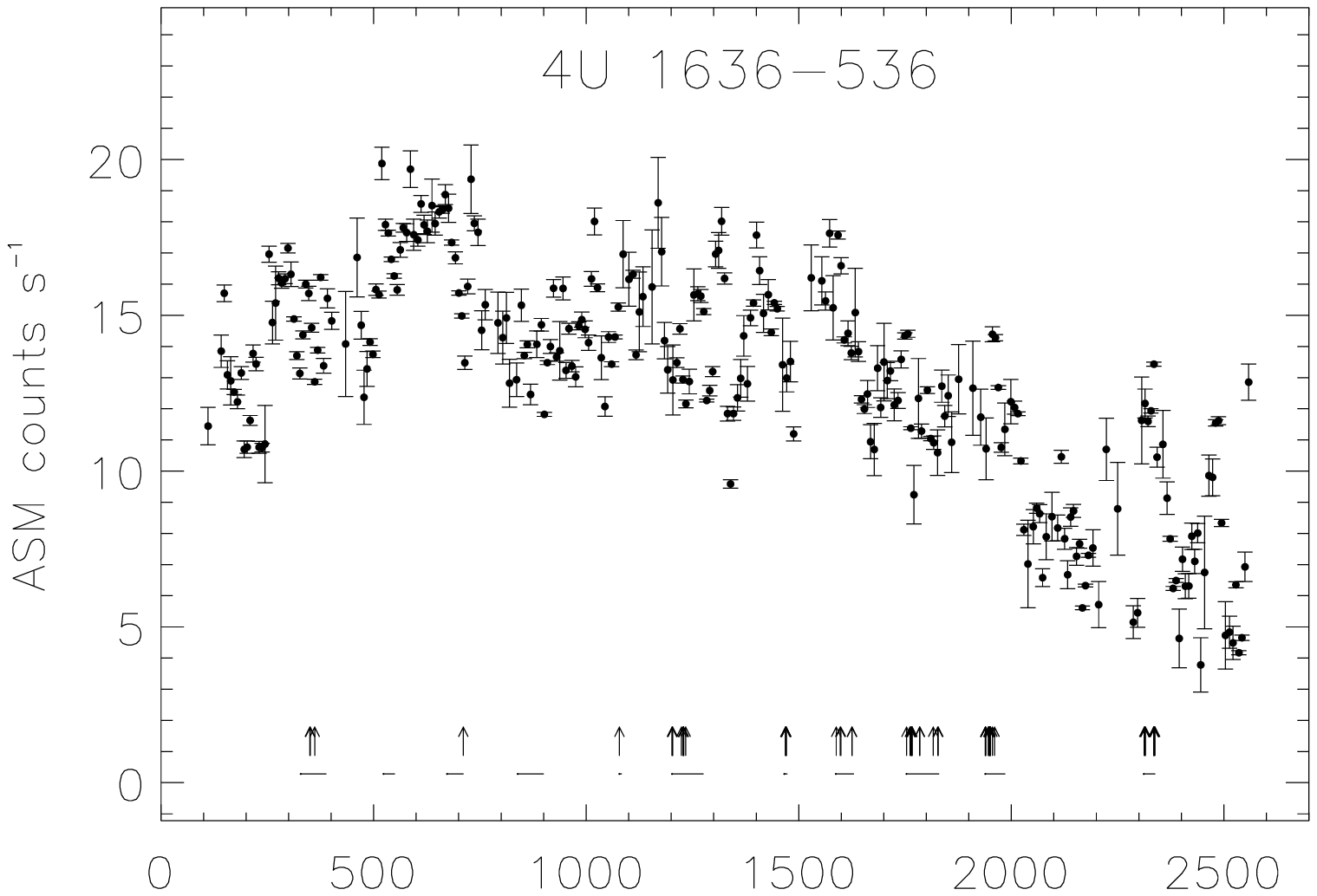,width=6.0cm,clip=t}}
\parbox[b]{6.0cm}{\psfig{figure=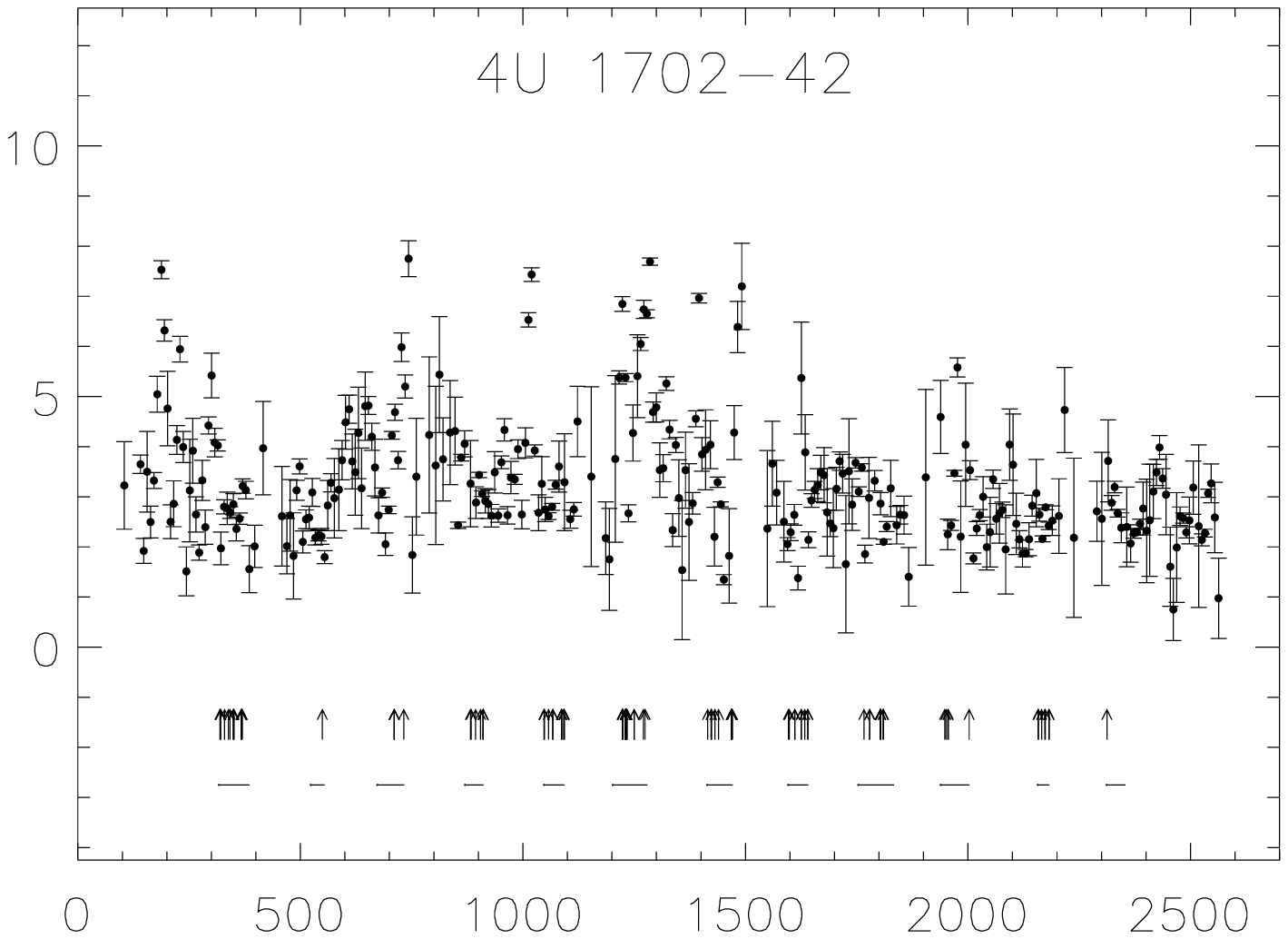,width=6.0cm,clip=t}}
\parbox[b]{6.0cm}{\psfig{figure=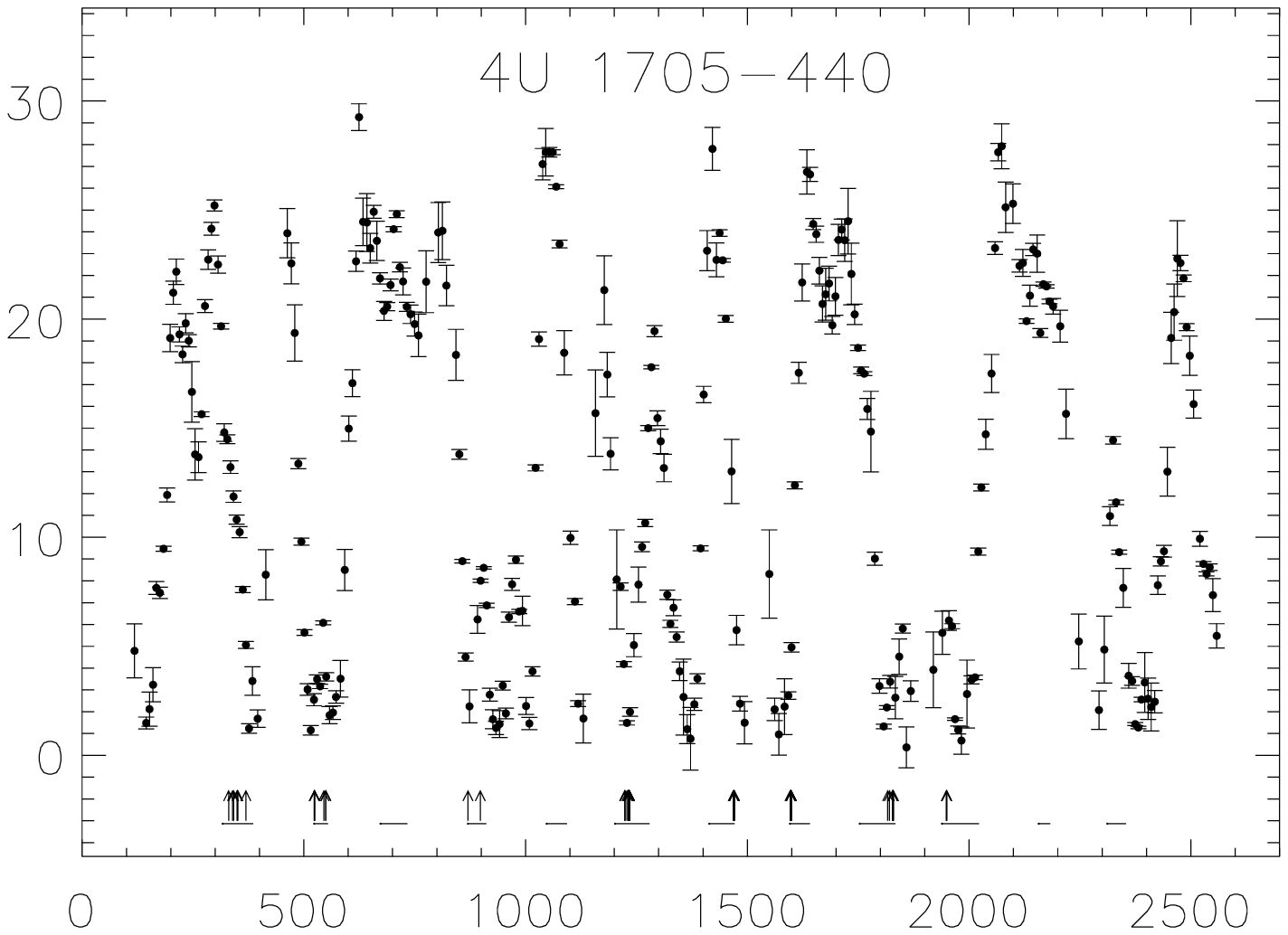,width=6.0cm,clip=t}}
\parbox[b]{6.0cm}{\psfig{figure=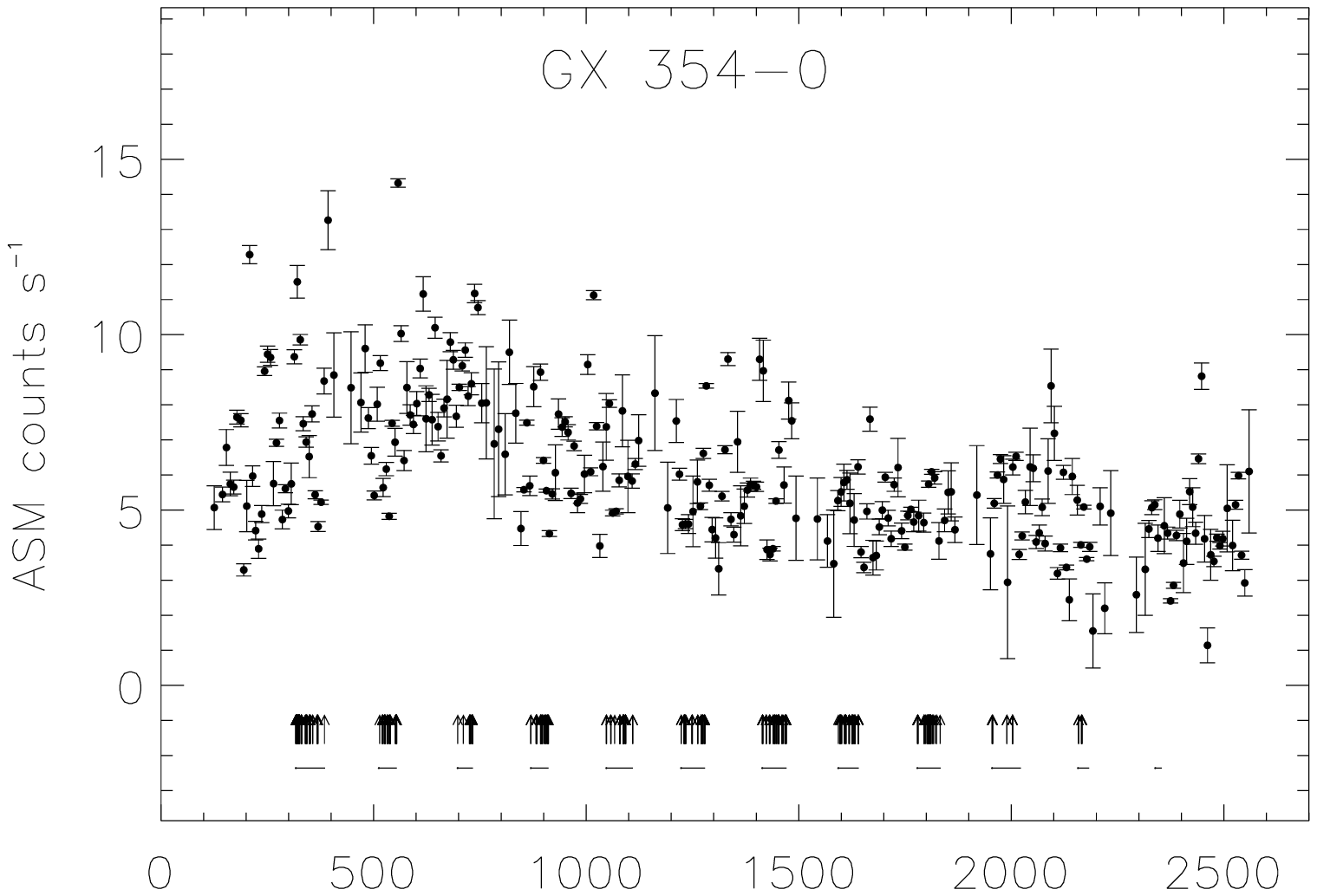,width=6.0cm,clip=t}}
\parbox[b]{6.0cm}{\psfig{figure=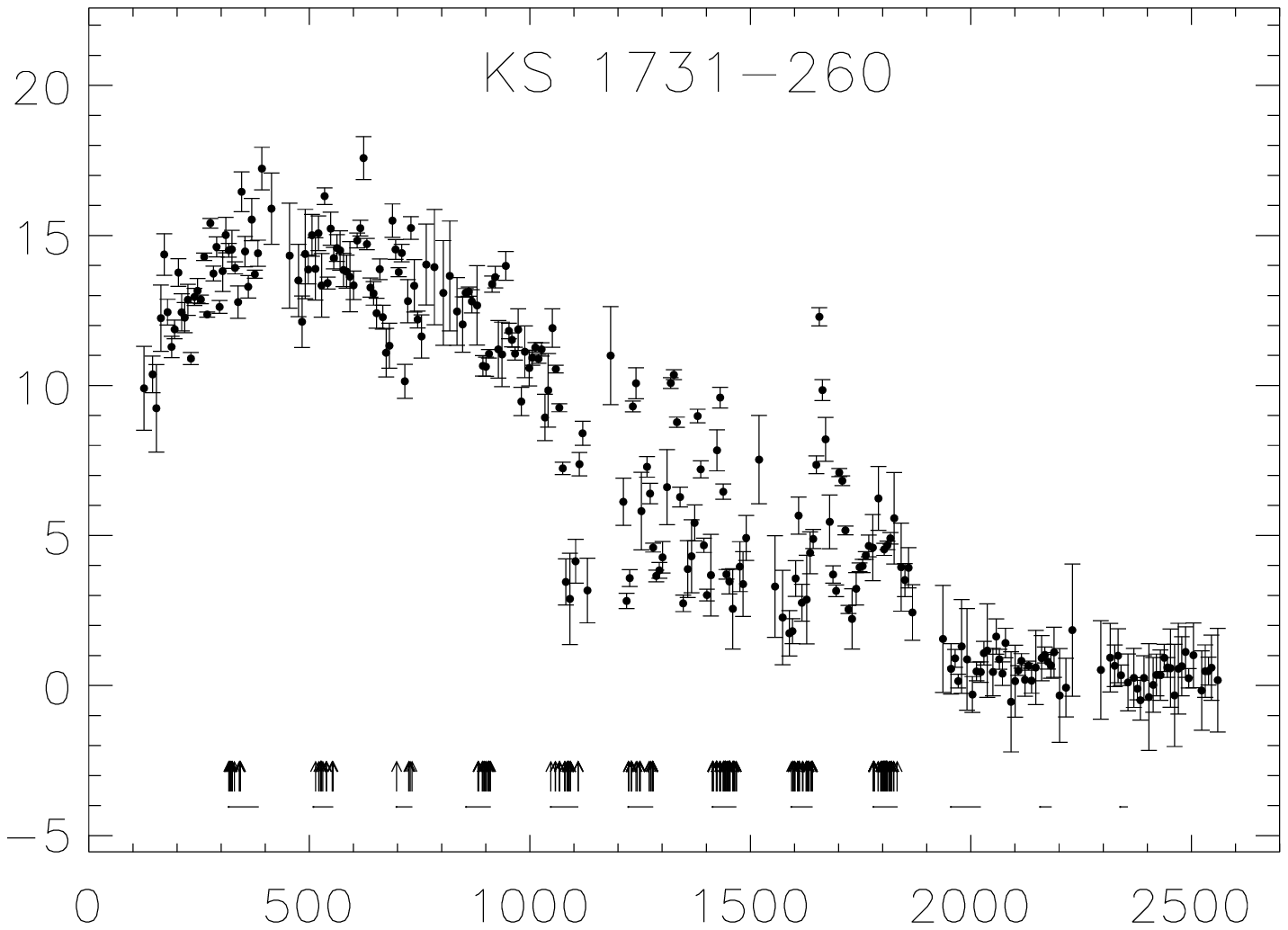,width=6.0cm,clip=t}}
\parbox[b]{6.0cm}{\psfig{figure=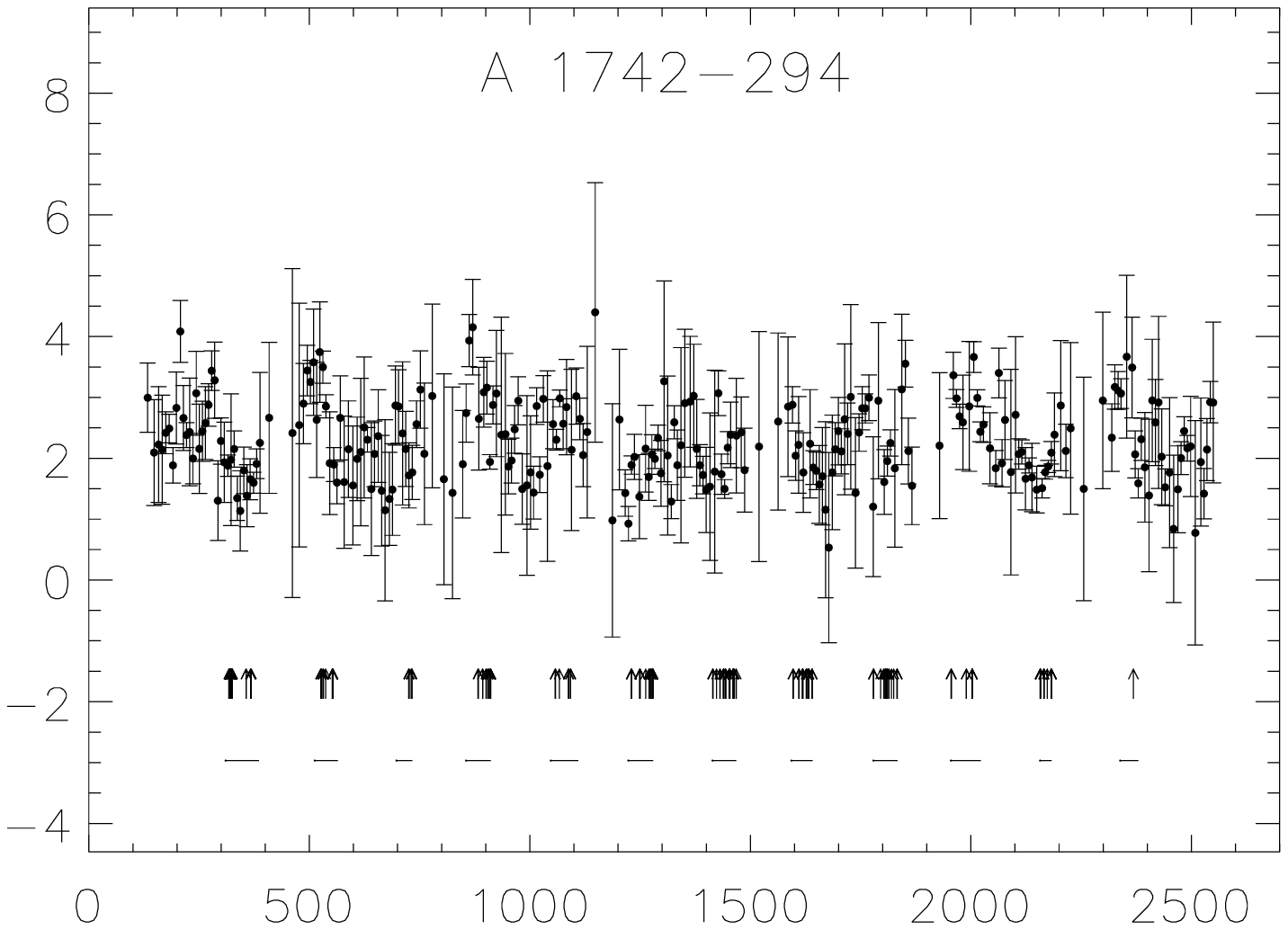,width=6.0cm,clip=t}}
\parbox[b]{6.0cm}{\psfig{figure=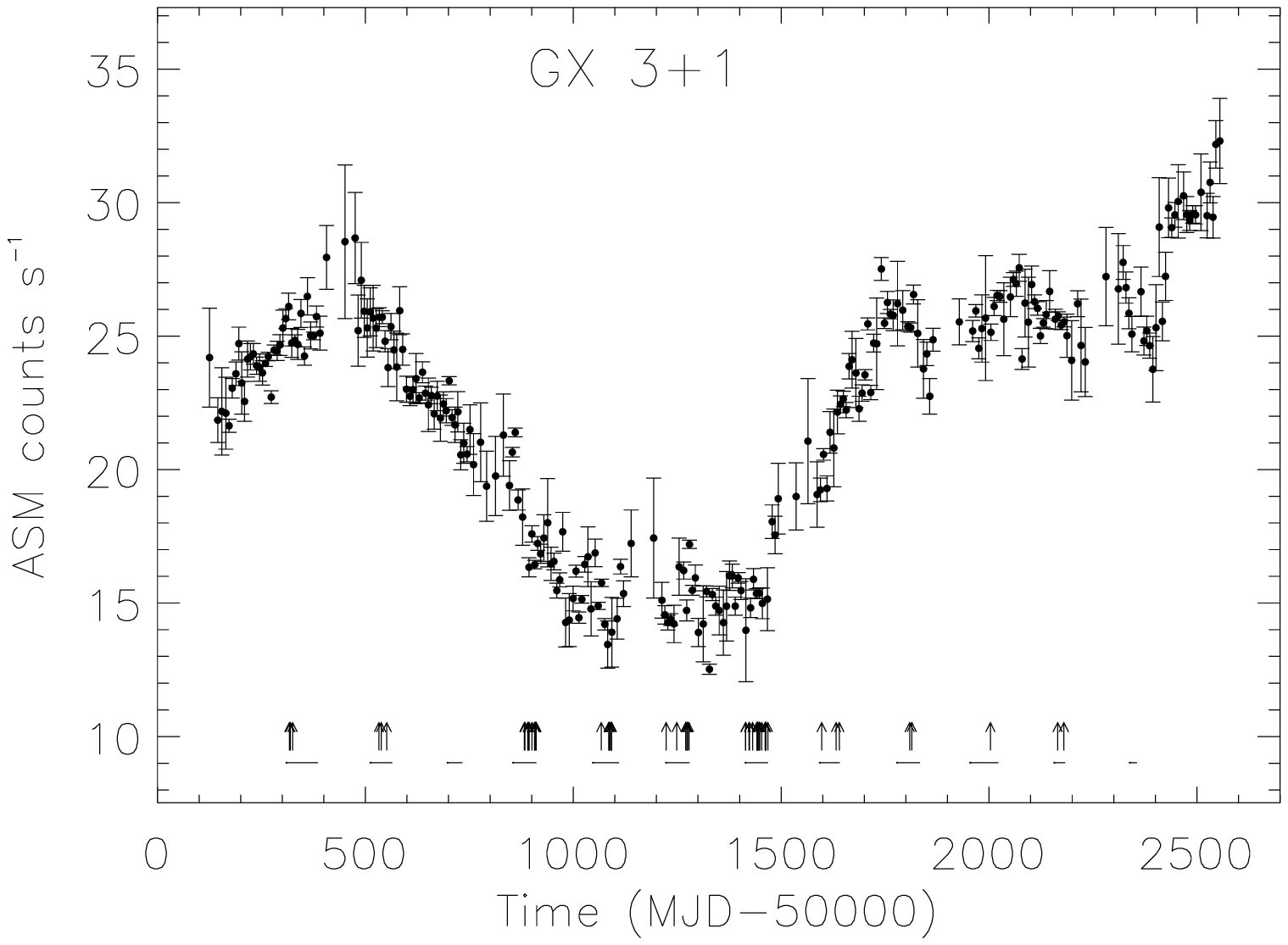,width=6.0cm,clip=t}}
\parbox[b]{6.0cm}{\psfig{figure=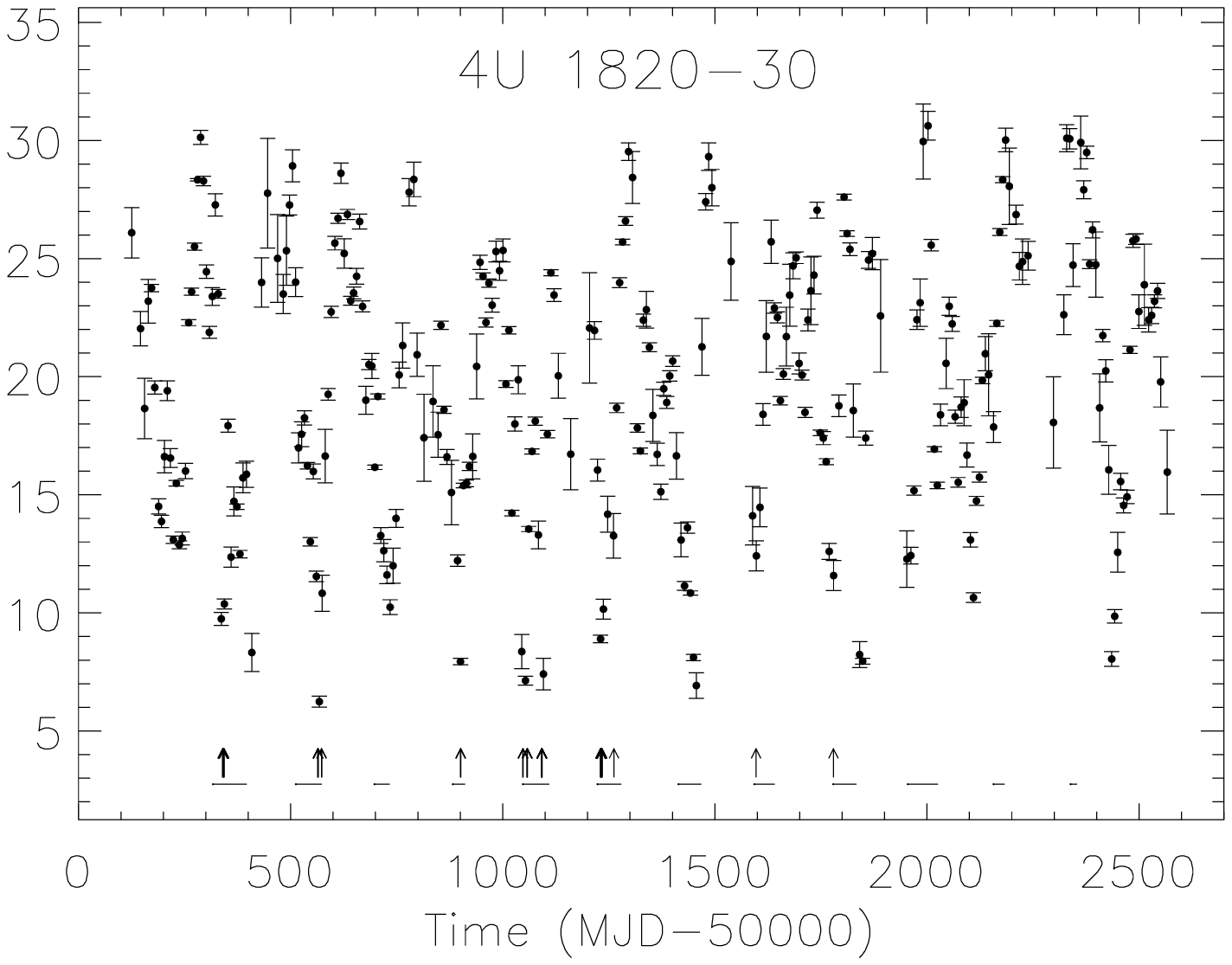,width=6.0cm,clip=t}}
\parbox[b]{6.0cm}{\psfig{figure=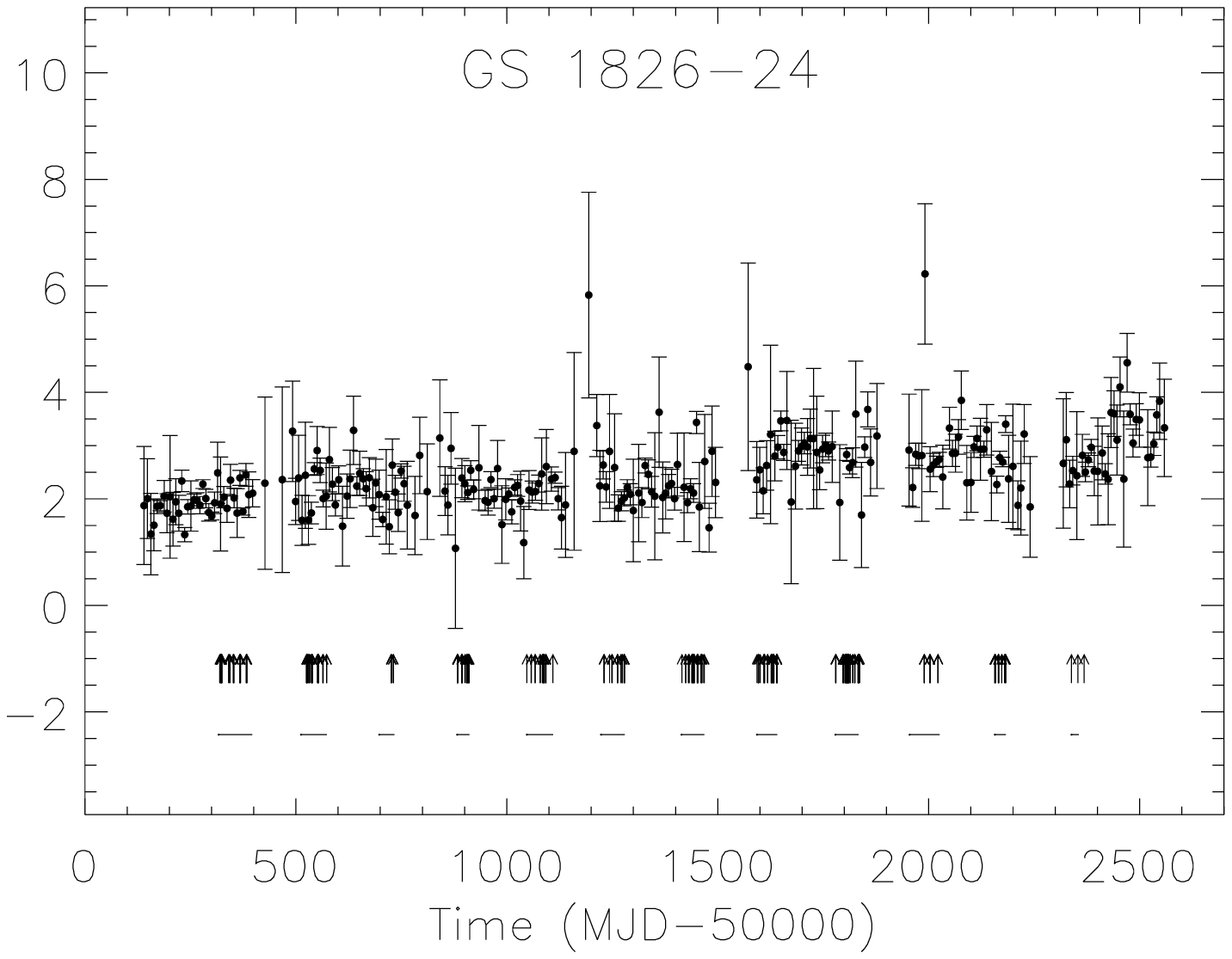,width=6.0cm,clip=t}}
\caption{ASM/RXTE lightcurves of 9 of the most frequent X-ray 
bursters in the WFC database. Each bin is a one week average.
Below the lightcurve the WFC observations on these sources are
indicated with horizontal bars. The arrows just above the horizontal
bars indicate the times of type\,I bursts. 
\label{asmcurves}}
\end{figure*}

We also searched the lightcurve of all photons detected on the whole
detector (``detector lightcurve'') for X-ray bursts from sources whose
persistent flux is below the detection limit of the WFC (see e.g.,
Cornelisse et~al. 2002); note that this limit becomes worse toward
the edge of the field of view.  Detector lightcurves are created with
a resolution of 1 s. A running average of 50 bins is calculated and if
at maximum 24 successive bins are $>$4\% above the average the first
bin is labeled as a candidate burst. For all potential new bursts we
cross-correlate the detector image again with the coded mask but only
for the burst time interval. In this way a point source from which the
burst originated may be identified. The sensitivity in this procedure
is at most a factor of 2 worse than in the above mentioned procedure.

\section{Results}

\subsection{Global burst behavior}

A total of 1823 bursts have been detected from 37 sources in the
Galactic center region, when not taking into account bursts from the
Rapid Burster (MXB\,1730$-$333) and the Bursting Pulsar
(GRO\,J1744$-$29). For each burst we determined the exponential decay
time in the total energy band (2-28 keV), the peak flux and average
persistent flux over the observation in WFC cts s$^{-1}$ cm$^{-2}$.
In Table\,\ref{top} we give an overview of the nine most frequent
bursters in the Galactic center region. All these sources are known
X-ray bursters and have been studied in the past.  The number of
type\,I bursts detected with the WFC in other burst sources is
too small for a meaningful statistical analysis.

In Fig.\,\ref{asmcurves} we show RXTE All Sky Monitor (RXTE/ASM;
Levine et~al. 1996) lightcurves of the sources listed in
Table\,\ref{top}.  Most lightcurves show a smooth variation and no
large fluctuations on a timescale of weeks. On the timescale of years
a variation by a factor of roughly 50\% is often present in these
sources, and an apparent increase in burst rate (indicated by arrows)
when the source becomes fainter. In contrast, the lightcurves of
4U\,1705$-$440 and 4U\,1820$-$30 show strong variations, and bursts
are only observed when the flux is low.  Note that bursts would have
been easily detected at the highest observed persistent flux levels of
all sources, because those levels are presumed to be still
significantly below the Eddington limit.

\begin{figure*}[t]

\vspace*{4.5cm}
\parbox[t]{6.0cm}{\epsfig{figure=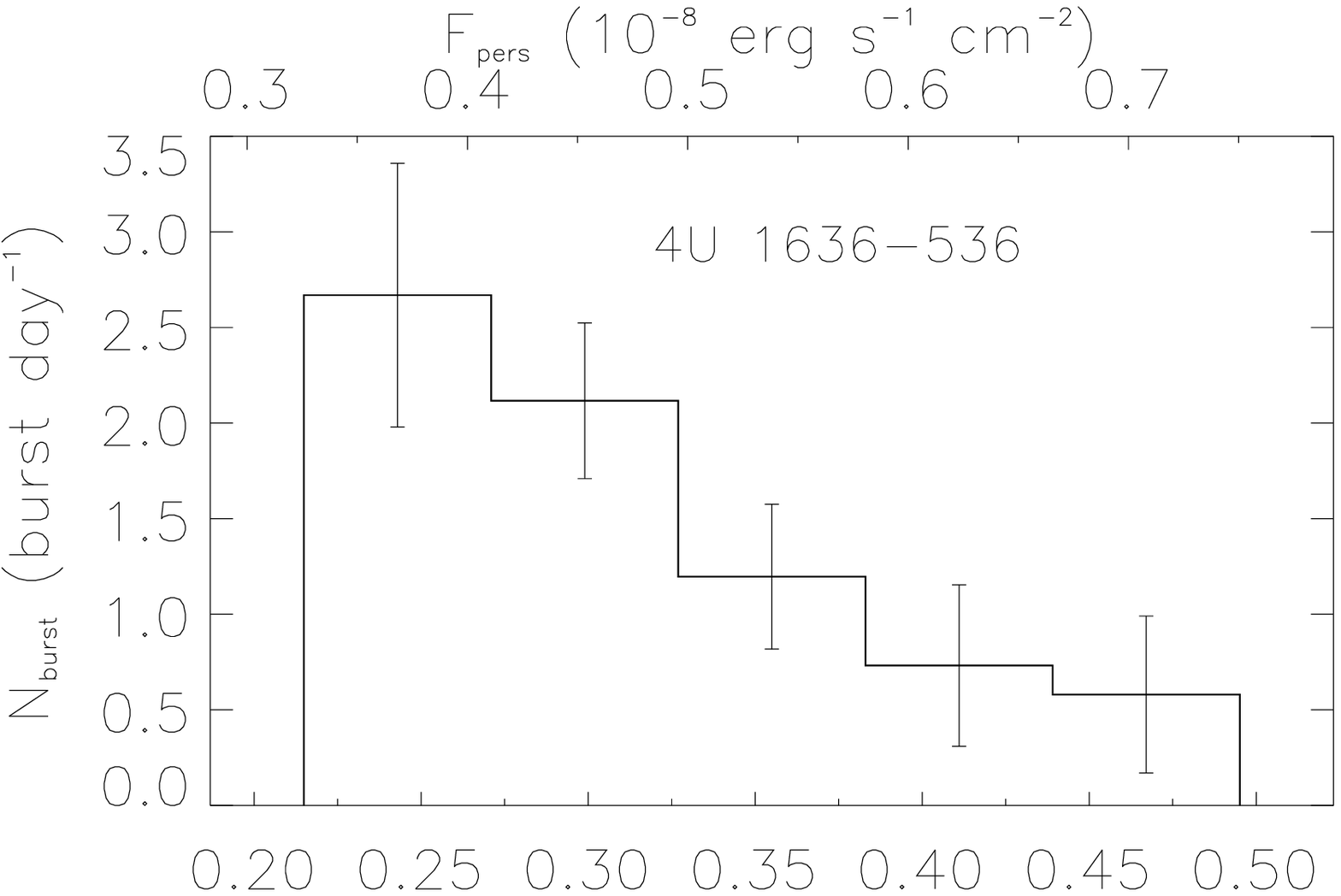,width=6.0cm,bburx=470,clip=t}}
\parbox[t]{6.0cm}{\epsfig{figure=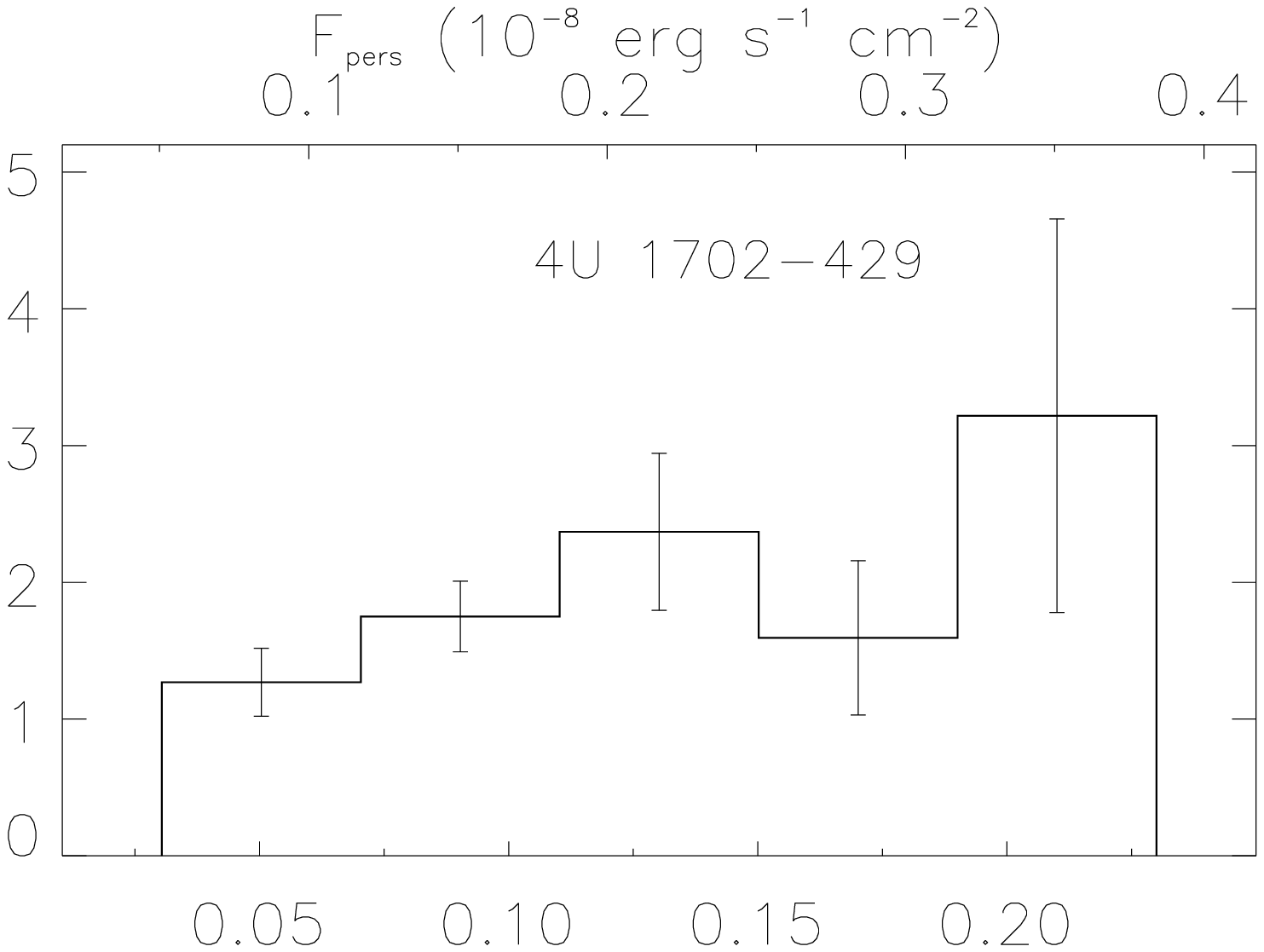,width=6.0cm,bburx=470,clip=t}}
\parbox[t]{6.0cm}{\epsfig{figure=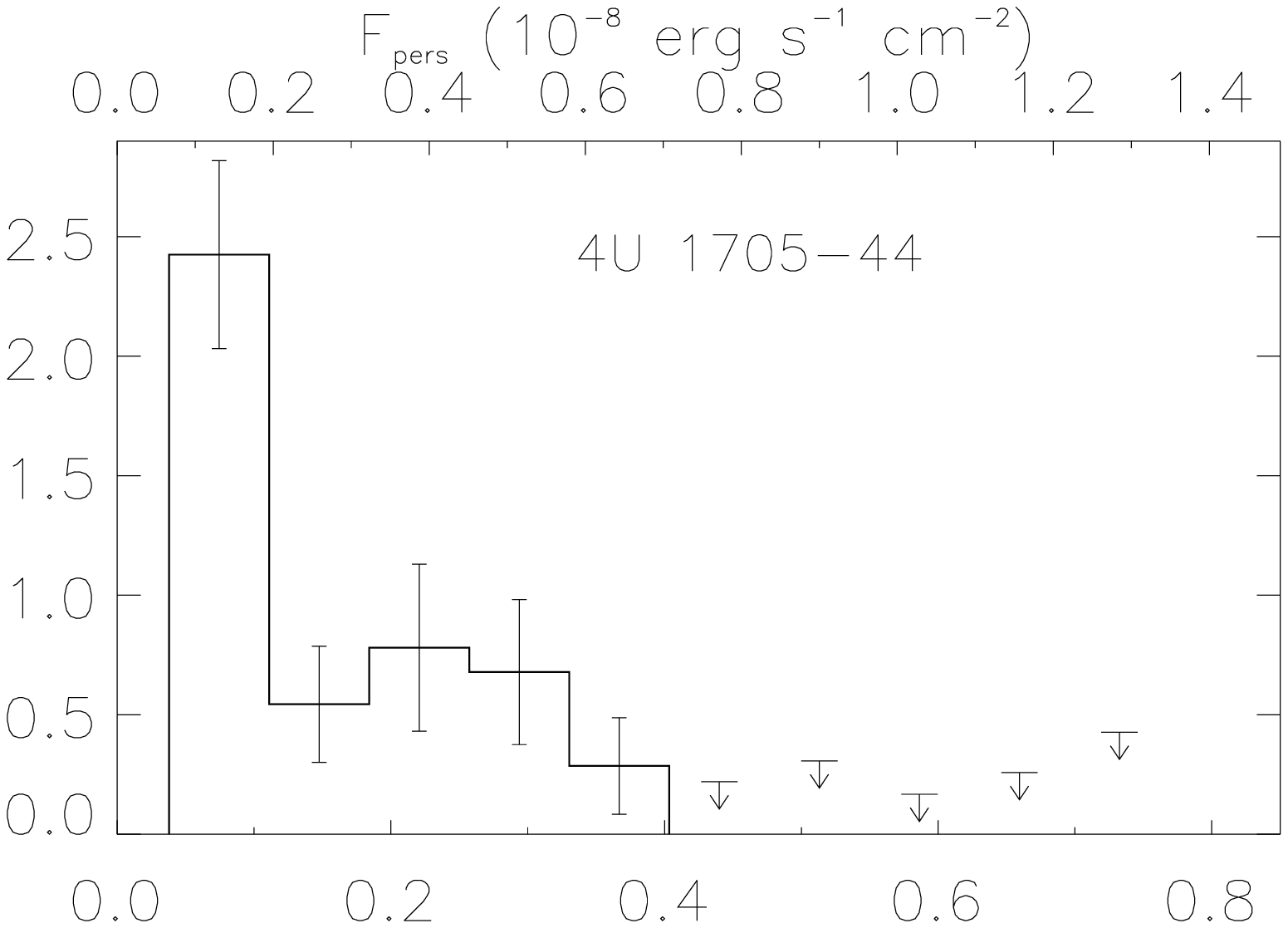,width=6.0cm,bburx=470,clip=t}}

\vspace*{3.8cm}
\parbox[b]{6.0cm}{\epsfig{figure=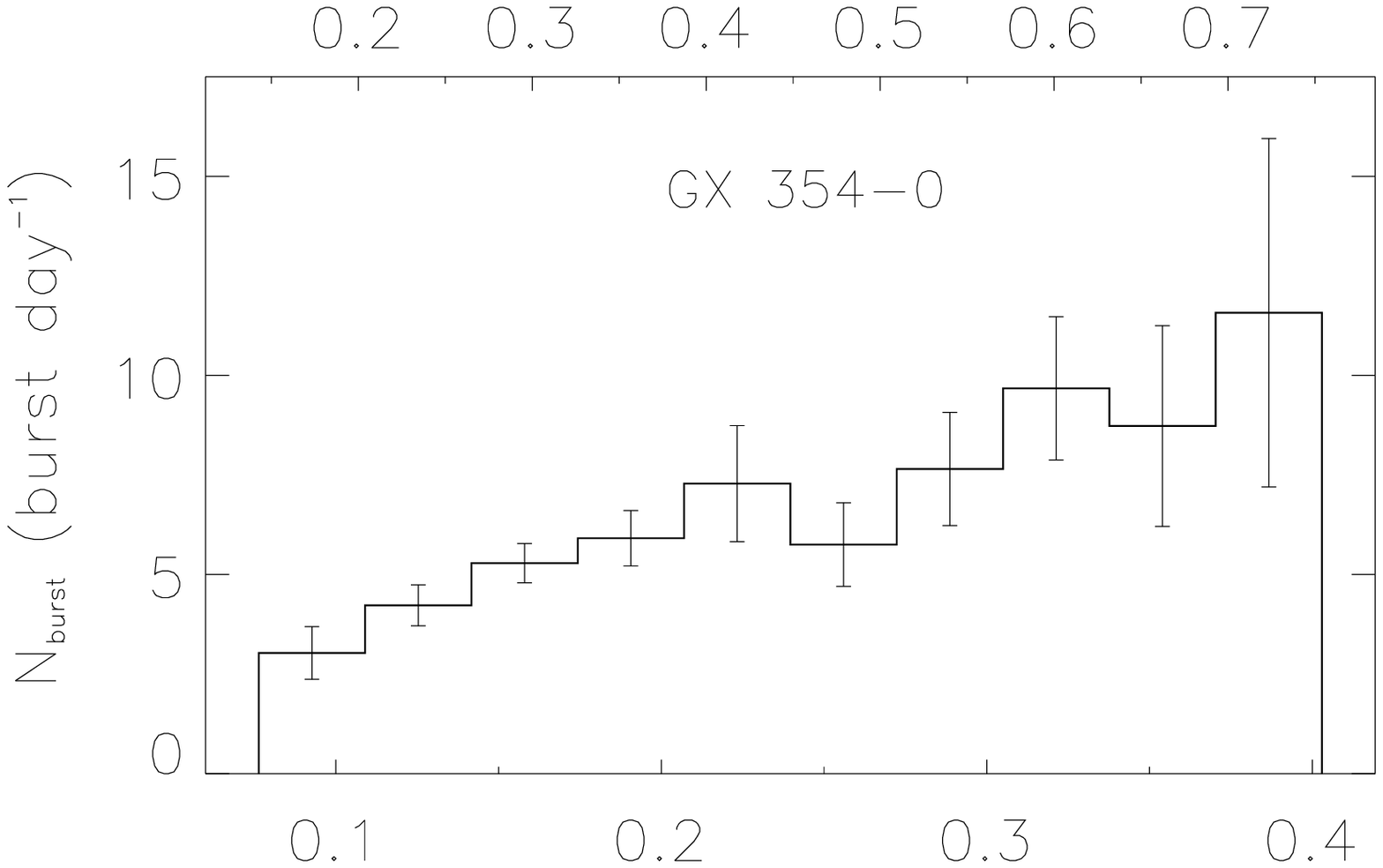,width=6.0cm,bburx=470,clip=t}}
\parbox[b]{6.0cm}{\epsfig{figure=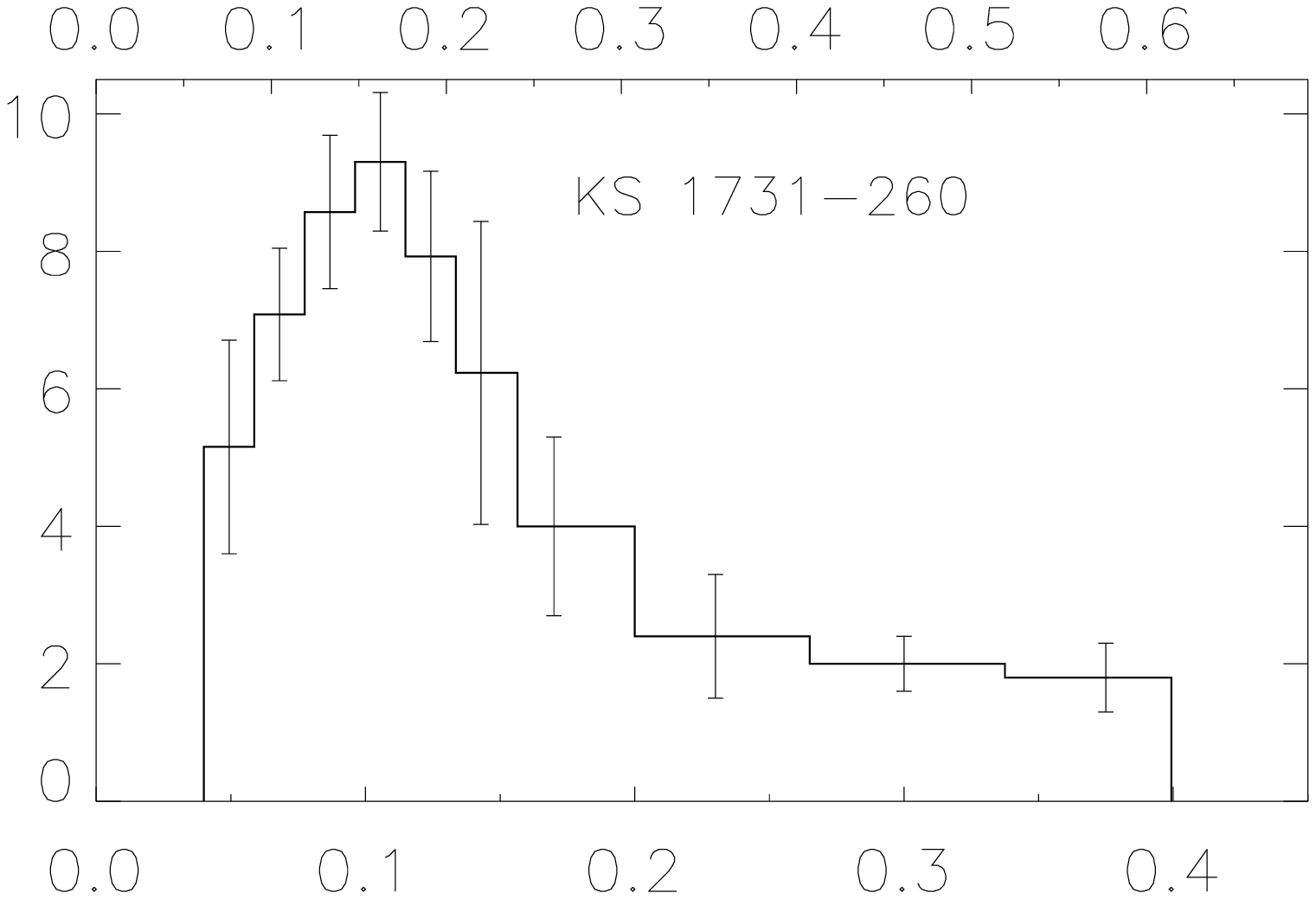,width=6.0cm,bburx=470,clip=t}}
\parbox[b]{6.0cm}{\epsfig{figure=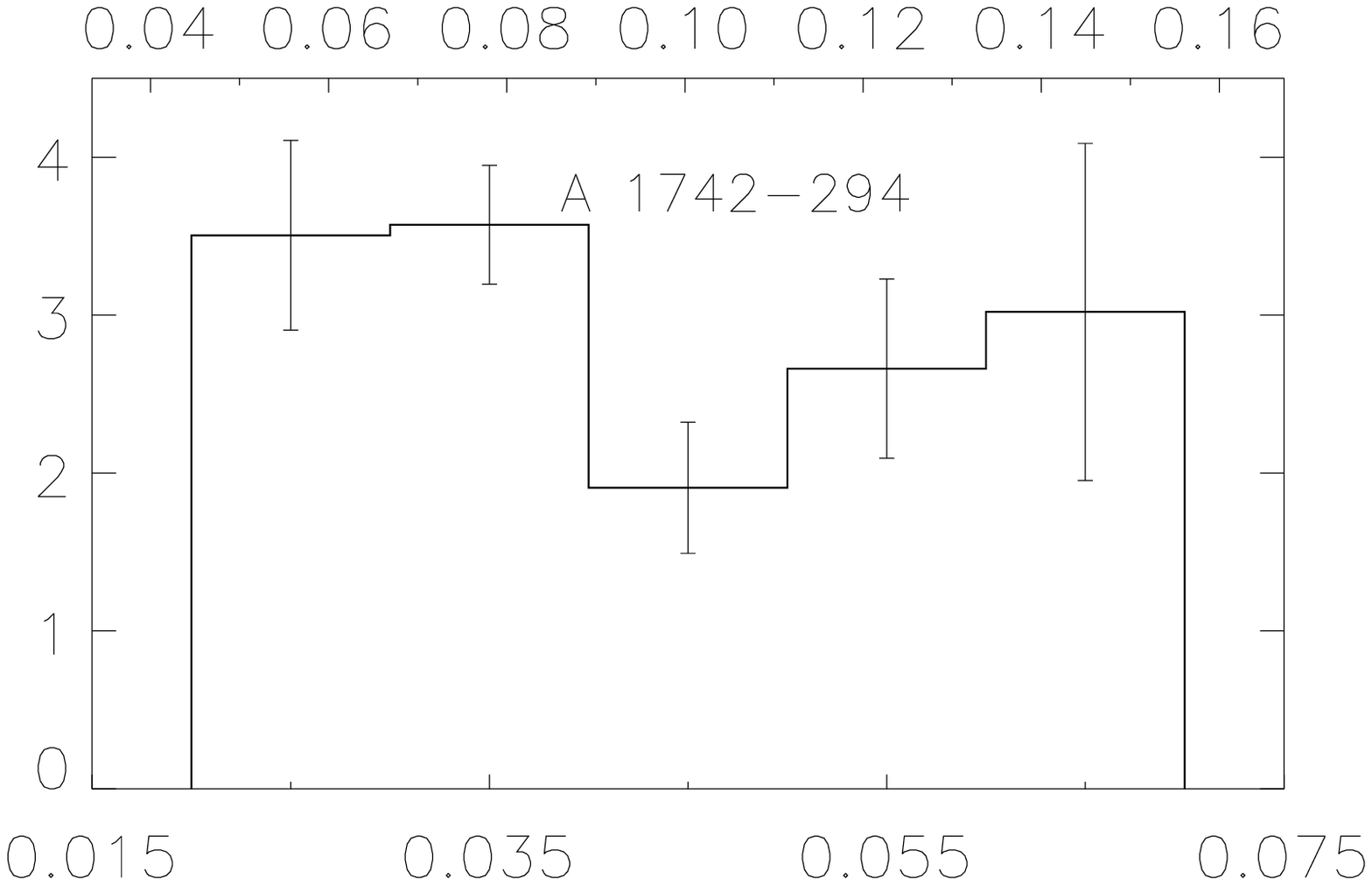,width=6.0cm,bburx=470,clip=t}}

\vspace*{3.8cm}
\parbox[t]{6.0cm}{\epsfig{figure=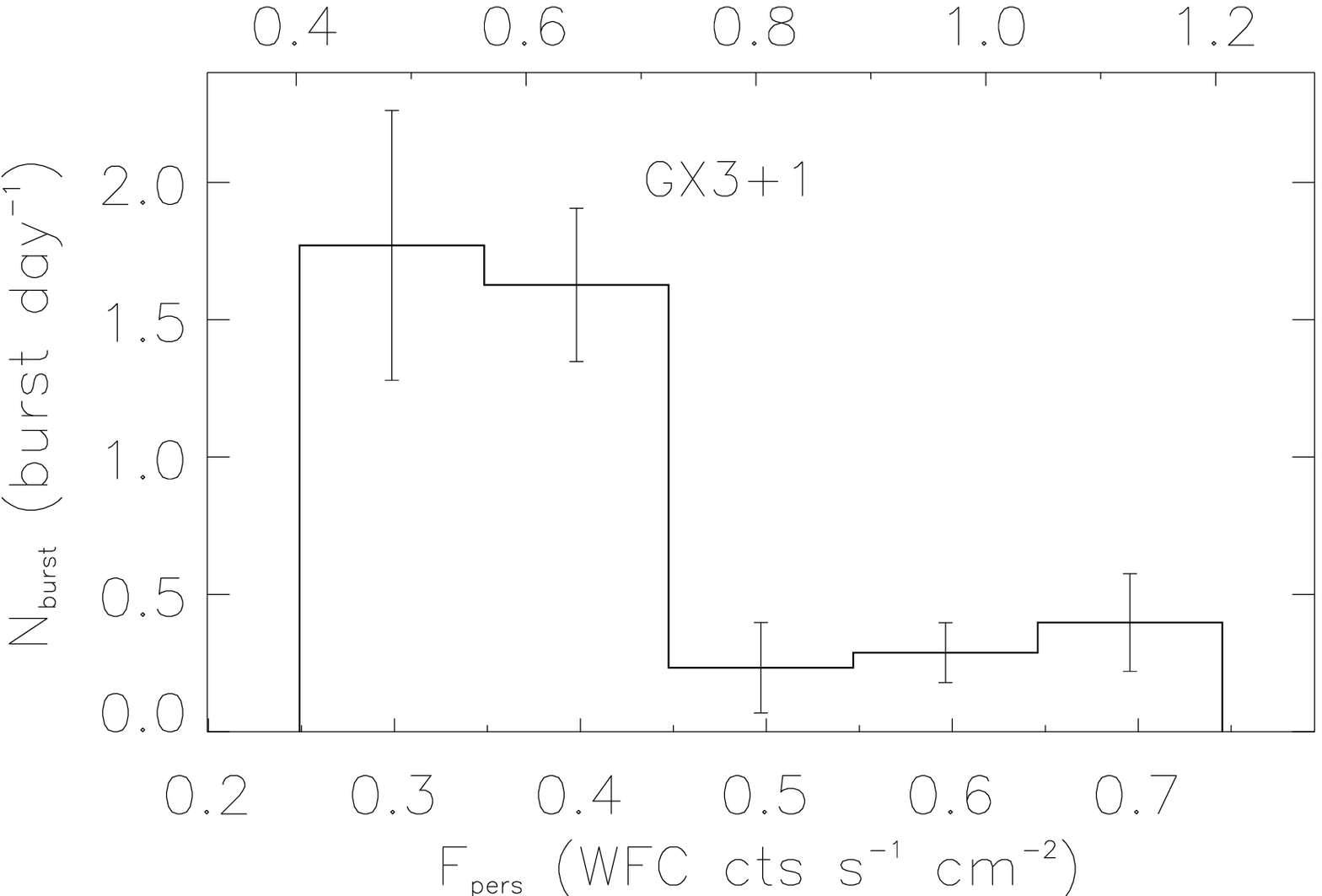,width=6.0cm,bburx=470,clip=t}}
\parbox[t]{6.0cm}{\epsfig{figure=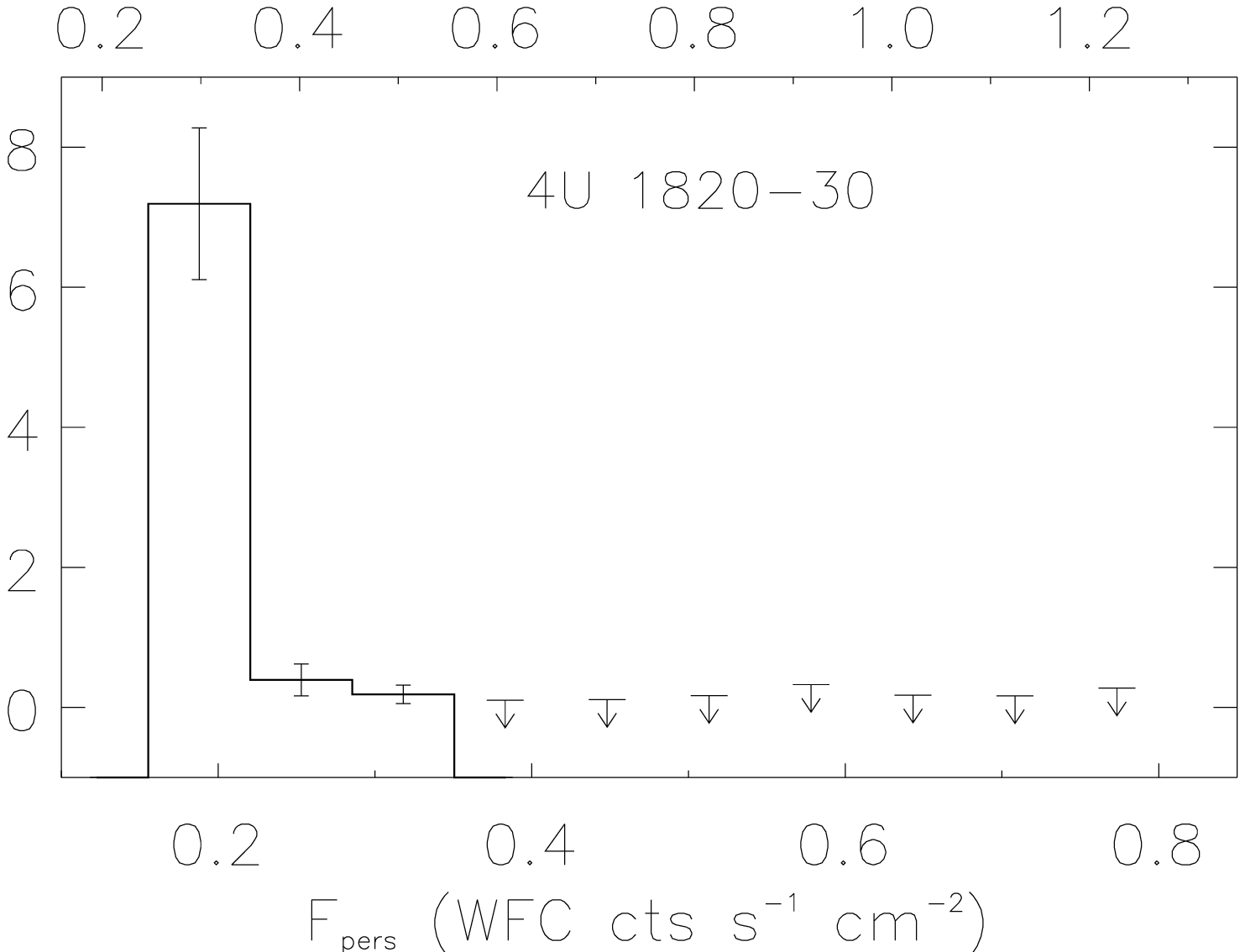,width=6.0cm,bburx=470,clip=t}}
\parbox[t]{6.0cm}{\epsfig{figure=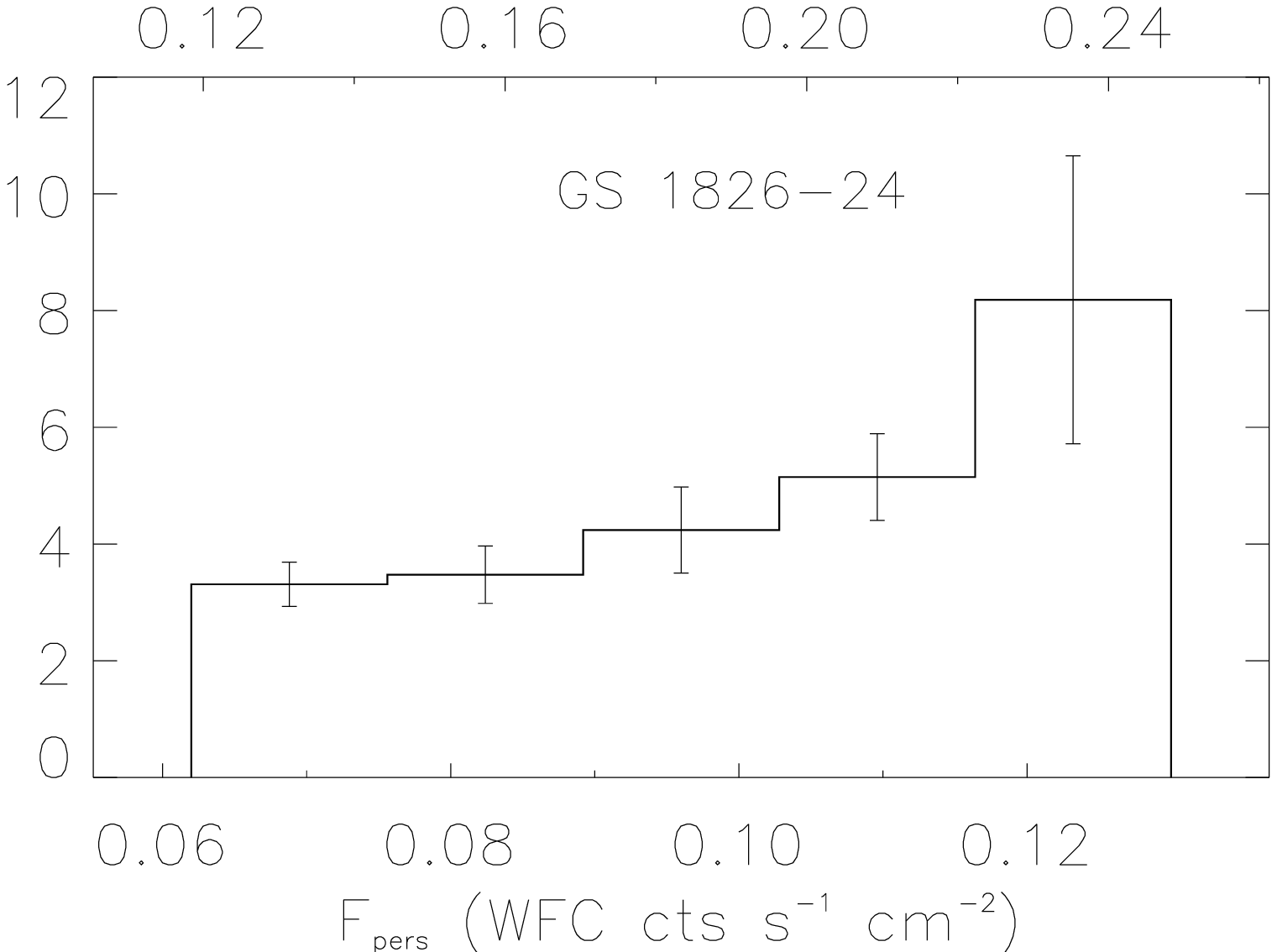,width=6.0cm,bburx=470,clip=t}}
\caption{Burst rate, N$_{\rm burst}$, as a function of the observed
count rate, F$_{\rm pers}$ (WFC cts s$^{-1}$ cm$^{-2}$; bottom axis)
and the persistent flux, F$_{\rm pers}$ (10$^{-8}$ erg s$^{-1}$
cm$^{-2}$; top axis) for the nine frequent X-ray bursters in the
galactic center region. The photon and energy flux are for a bandpass
of 2-28 keV.
\label{burstflux}}
\end{figure*}

To study the burst rate as a function of persistent flux we assigned
to each burst the average WFC-measured flux over the complete
observation in which the burst occurred. We also checked the average
persistent flux in a 5-minute time interval prior to each burst, but
the difference with the aforementioned flux is within the errors. The
persistent flux range was divided in 5 or 10 intervals of equal
size. Only the flux range of KS\,1731$-$260, the sole transient source
in our sample, was divided in 10 bins with different bin sizes.  For
each flux interval we determined the total exposure time and the
number of bursts; the latter with an error equal to the square root of
this number (if no bursts are within an interval upper limits of 68\%
confidence were estimated). In Fig.\,\ref{burstflux} we show for each
source the burst rate as a function of observed photon flux (bottom
axis) and the derived energy flux (top axis).  To get an indication of
the conversion of photon flux to energy flux we generated for each
source a spectrum for each campaign.  We fitted the spectrum, assuming
an absorbed thermal bremsstrahlung model, and we derived a mean
conversion factor over all campaigns (corrected for absorption), see
Table\,\ref{top}. The spread in the conversion factors over all
campaigns is about 10\% for each source. In the WFC passband the
assumption of a bremsstrahlung spectrum is good enough to derive
fluxes; fluxes derived using other models give comparable results.

We notice in Fig.\,\ref{burstflux} that at the lowest flux levels
KS\,1731$-$260 shows an increase in burst rate with increasing
persistent flux.  When KS\,1731$-$26 reaches $1.7\times10^{-9}$
$\ergcms$, the burst rate drops by a factor of 5 and at higher flux
levels the burst rate slowly decreases. 4U\,1702$-$429, GX\,354$-$0
and GS\,1826$-$24 all show only an increasing burst rate, while
4U\,1636$-$536 is the only source that shows a decreasing burst rate.
4U\,1705$-$44, GX\,3$+$1 and 4U\,1820$-$30 show a drop by a factor of
$\simeq$5 in burst rate over a small range of persistent flux.
A\,1742$-$294 is the only source for which no clear trend is visible, and
the burst rate stays constant over the total observed flux range.
This source traces out the lowest fluxes within our sample.

\begin{figure*}
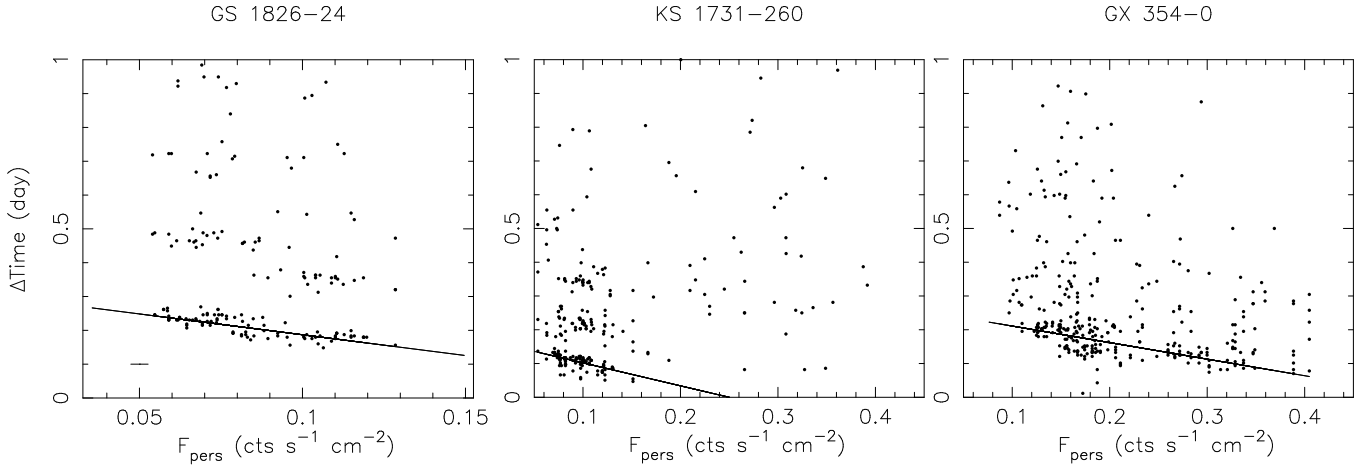

\parbox[b]{6.5cm}{\psfig{figure=H4336F19.eps,angle=-90,width=6.4cm}}
\parbox[b]{5.6cm}{\psfig{figure=H4336F20.eps,angle=-90,width=5.6cm}}
\parbox[b]{5.6cm}{\psfig{figure=H4336F21.eps,angle=-90,width=5.6cm}}
\caption{The wait time, $\Delta$Time, as a function of persistent
flux, F$_{\rm pers}$, for the sources GS\,1826$-$24 (left panel) and
KS\,1731$-$260 (middle panel) and GX\,354$-$0 (right panel). A typical
error in the flux for GS\,1826$-$24 is indicated at the bottom left,
while the error in the flux for the other two sources is as large as a
dot. The error in the wait time for all sources is much smaller than
the size of a dot. A best linear fit is drawn for the bursts for which
the previous burst is not missed.
\label{waiting}}
\end{figure*}

\subsection{Wait times}

Several sources are known to show quasi-periodic burst recurrence
times during certain periods.  The best example is GS\,1826$-$24,
in 1996-1997 it exhibited a burst every $\simeq$6 hours, and the
burst wait times were constant within a few minutes for long periods
of time (Ubertini et~al. 1999; Cocchi et~al. 2001b).

In the left panel of Fig.\,\ref{waiting} we plotted the wait time as a
function of the persistent flux for GS\,1826$-$24. Most wait times
appear to follow a straight line at the bottom of the figure. A second
linear trend can clearly be distinguished above this line (and even
two more above that).  Given the fact that BeppoSAX has a 96-minute
low-earth orbit, it is probable that bursts are missed during data
gaps and that multiples of the burst wait times are observed. We
checked bursts with long wait times within one observation and find
that for all of them the previous burst may have occurred
during an earth occultation or South Atlantic Anomaly passage.  From
Fig.\,\ref{waiting} we see that the wait time between the bursts
decreases linearly with increasing persistent flux. We performed a
least-squares fit on the bursts where the previous one is {\it not}
missed to the function:
\begin{math}
\Delta t=AF_{\rm pers}+B.
\end{math} 
The results for the parameters are given in Table\,\ref{timefit}.
 Assuming a wait time that is two times longer (i.e., doubling the
 numbers derived above) gives a good description of the trend followed
 by the bursts forming the second line from the bottom. This shows
 again that for these points the previous burst is missed.  Note that
 formally the fit is not acceptable ($\chi^2_\nu=5.7$, 92 d.o.f.), but
 the general trend is clearly visible. This means that there are
 significant fluctuations in the wait time around the average
 relation.

From theory a linear relation between the burst rate (inverse of the
wait time) and the persistent flux is expected, and no bursts are
expected anymore when $F_{\rm pers}$=0 $\ergcms$ (i.e. no accretion).
We therefore tried to fit the relation:
\begin{math}
\Delta t=C/F_{\rm pers}.
\end{math}
The result of the fit of parameter C is given in Table\,\ref{timefit}.

Given the large number of bursts, we have also investigated the
relation between the wait time and persistent flux for KS\,1731$-$260
and GX\,354$-$0. In the middle panel of Fig.\,\ref{waiting} we show
the results for KS\,1731$-$260, and notice the strong suggestion of a
linear dependency.  However, this only applies to persistent flux
levels below 0.14 WFC\,cts\,cm$^{-2}$\,s$^{-1}$. At higher persistent
flux the wait time between bursts becomes apparently random. We fitted
the relations as given above for the bursts with a persistent flux
below 0.14 WFC\,cts\,cm$^{-2}$\,s$^{-1}$ and where we expect that the
previous burst is not missed. The best fit parameters are given in
Table\,\ref{timefit}.

Also for GX\,354$-$0 there vaguely appears to be a linear relation
between the persistent flux and the wait time (right panel
Fig.\,\ref{waiting}).  However, the scatter is significantly larger
than in the previous two cases, making a clear distinction between the
different multiples of the wait time very difficult. Therefore, an
iterative process was used to search for bursts where we expect that
the previous one is not missed.  We simultaneously fitted a straight
line (the single wait time line) plus several lines at multiples of
the wait time. The bursts closest to the single wait time line were
attributed to this line and used for a least-square fit to get a
better estimate. This process was continued until a best fit was
found. The bursts attributed to the single wait time line were also
used to fit the relation:
\begin{math}
\Delta t=C/F_{\rm pers}.
\end{math}
The results are summarized in Table\,\ref{timefit}.

\begin{table}
\caption{Best fit parameters for a relation where the
wait time is proportional (A and B) or inversely proportional
(C) to the persistent flux. Due to the large $\chi^2_\nu$ the 
formal errors on the parameters have no meaning.  
\label{timefit}}
\begin{tabular}{lccc}
\hline
\hline
parameter & GS\,1826$-$24 & KS\,1731$-$260 & GX\,354$-$0\\ 
\hline
 A & -1.23$\pm$0.10 & -0.68$\pm$0.09 & -0.49$\pm$0.07\\
 B &  0.31$\pm$0.01 &  0.17$\pm$0.01 & 0.26$\pm$0.01\\
$\chi^2_\nu$ (d.o.f.) & 5.7 (92) & 20 (100) & 266 (202)\\
\hline
 C  &  0.017$\pm$0.001 & 0.009$\pm$0.001 & 0.029$\pm$0.001\\
$\chi^2_\nu$ (d.o.f.) & 6.2 (93) & 38 (101) & 234 (203)\\
\hline
\end{tabular}
\end{table}

\begin{figure*}[t]
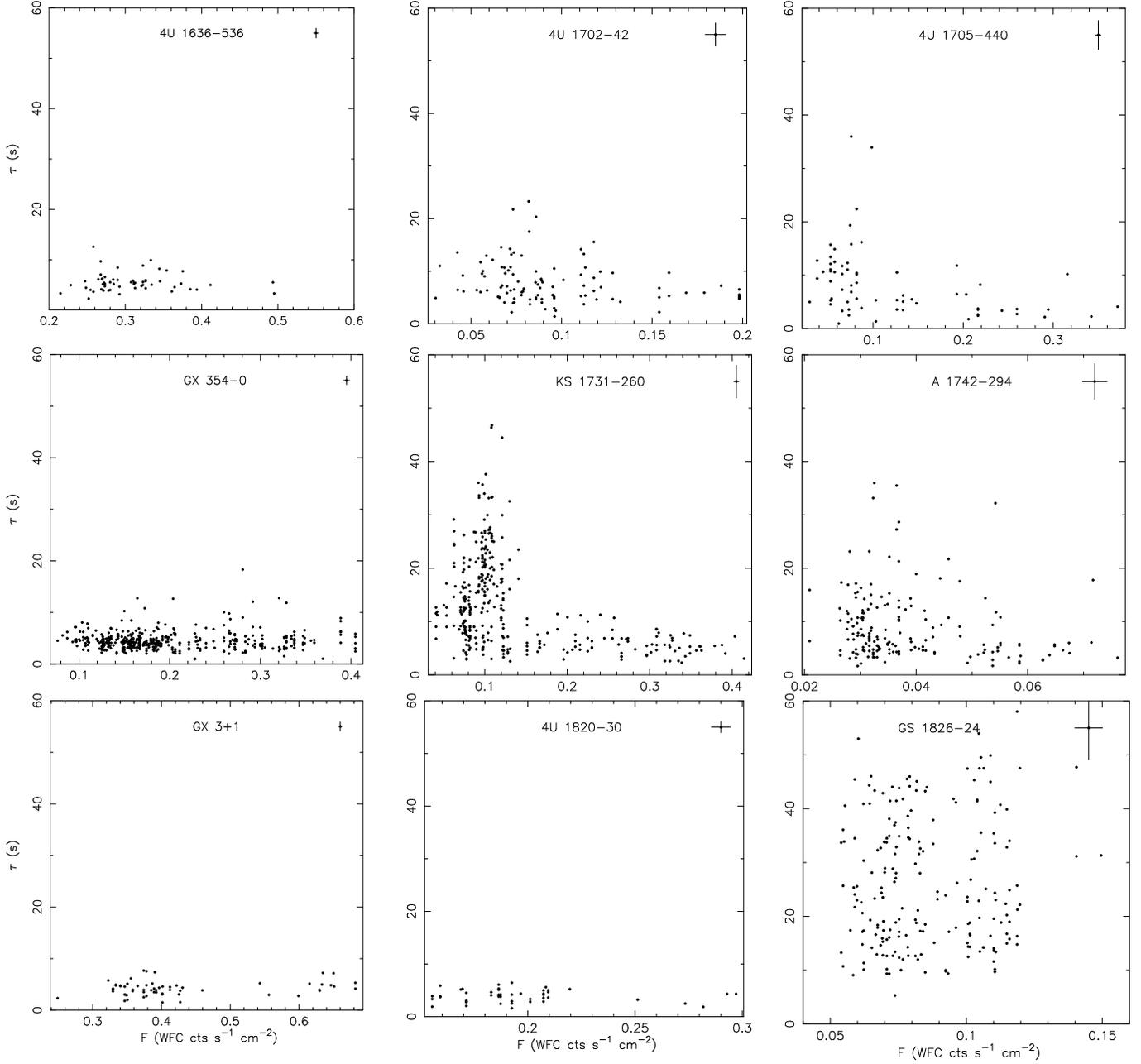

\parbox[b]{6.0cm}{\psfig{figure=H4336F22.eps,angle=-90,width=5.75cm}}
\parbox[b]{5.5cm}{\psfig{figure=H4336F23.eps,angle=-90,width=5.5cm}}
\parbox[b]{5.5cm}{\psfig{figure=H4336F24.eps,angle=-90,width=5.5cm}}
\parbox[b]{6.0cm}{\psfig{figure=H4336F25.eps,angle=-90,width=5.75cm}}
\parbox[b]{5.5cm}{\psfig{figure=H4336F26.eps,angle=-90,width=5.5cm}}
\parbox[b]{5.5cm}{\psfig{figure=H4336F27.eps,angle=-90,width=5.5cm}}
\parbox[b]{6.0cm}{\psfig{figure=H4336F28.eps,angle=-90,width=5.75cm}}
\hspace*{0.3cm}
\parbox[b]{6.0cm}{\psfig{figure=H4336F29.eps,angle=-90,width=5.55cm}}
\parbox[b]{6.0cm}{\psfig{figure=H4336F30.eps,angle=-90,width=5.55cm}}
\caption{The exponential decay time, $\tau$, as a function of persistent
flux, F, for the nine most frequent burst sources observed with
WFC. In the upper-right corner of each panel a typical error-bar
is shown.
\label{decay}}
\end{figure*}

For the other six sources the number of subsequent bursts with a wait
time of less than one day becomes very small, and the data does not
allow the verification of a linear relation. 

We converted the fit parameters as given in Table\,\ref{timefit} from
the observed flux to luminosities using the conversion factors and
distances as given in Table\,\ref{top}. For the parameter C the values
are $(2.5\pm1.5)\times10^{36}$, $(0.8\pm0.5)\times10^{36}$ and
($1.9\pm1.1)\times10^{36}$ $\ergs$ for GS\,1826$-$24, KS\,1731$-$260
and GX\,354$-$0, respectively (taking into account an error of 30\% in
the distance). Although the errors are very large, these slopes are
the same within their errors. Therefore, the burst rate may be a
unique function of the persistent flux.

\subsection{Decay times}

Another important burst parameter is the e-folding decay time. As
discussed in Sect.\,1, this diagnoses the composition
of the burst fuel. To derive the decay time for each burst
we generated lightcurves with a 1 s time resolution. A running average
of 5 s was used to determine the moment of the peak flux. The
persistent emission level and the decay time are then simultaneously
fitted with a constant and exponential, respectively. We took the bin
in which the peak flux was reached as the first data point. In
Fig.\,\ref{decay} we show the decay times as a function of persistent
emission for the nine sources. We here discuss GX\,354$-$0, KS\,1731$-$260
and GS\,1826$-$24 in more detail and compare them with the other
sources.

GX\,354$-$0 only shows bursts with decay times shorter than $\simeq$10
seconds at all flux levels.  The same applies to 4U\,1636$-$536,
GX\,3$+$1 and 4U\,1820$-$30.  GS\,1826$-$24 shows a large range of
decay times at all persistent flux levels, but almost no bursts below
10 seconds are observed (i.e., less than 5\% of all bursts). Of the nine
sources, this is the only one that shows this behavior.

Two trends can be observed for the decay times of KS\,1731$-$260. At
high persistent flux ($\gtap$0.14 WFC\,cts\,s$^{-1}$\,cm$^{-2}$) all
decay times are below 10 seconds, as for e.g. GX\,354$-$0.  At
lower fluxes the spread in decay times increases rapidly and most
bursts have decay times well above 10 s, as for GS\,1826$-$24.
However, in contrast to GS\,1826$-$24, still a significant fraction of
bursts show decay times below 10 s (about 30\%). The same behavior
is also suggested by the figures for 4U\,1702$-$429, 4U\,1705$-$44 and
A\,1742-294.

a spectral change instead of a change in persistent emission could be
the only indication of a change in mass accretion (van der Klis
et~al. 1990). This could explain the occurrence of both long and short
bursts at low persistent flux. We therefore investigated the low
persistent flux levels of KS\,1731$-$260 in a little more detail. At
MJD 51799.60 and MJD 51799.72 there were bursts with decay times of
4.7$\pm$0.1 s and 20.8$\pm$3.5 s, respectively. Due to the low flux
level of the source full resolution spectra do not have enough
statistics, and we resorted to the study of hardness ratios. The WFC
passband was divided in two channels from 2-6 keV and 6-28 keV, and
derived hardness ratios by dividing the count rates in the 6-28 keV
with the count rates in the 2-6 keV band. We found hardness ratios of
0.68$\pm$0.07 and 0.70$\pm$0.08 for the periods prior to the two
bursts, respectively.  We note that the average persistent flux stayed
constant at 0.111 and 0.115 WFC\,cts\,s$^{-1}$\,cm$^{-2}$ in these
periods. The 1$\sigma$ statistical fluctuations at one minute time
resolution are 24\% and 30\%, respectively. We conclude that no
significant changes occurred between the two bursts.

The transition from short bursts to long/short bursts in
KS\,1731$-$260 is rapid. Therefore, an observation (at MJD 51637)
where the persistent emission is at this transition was analyzed in
more detail. A spectrum was derived for this observation, assuming an
absorbed bremsstrahlung spectrum with a hydrogen absorption column of
$1.3\times10^{22}$ $\cmsq$ (Predehl \& Schmitt 1995). A temperature of
21.4$\pm$3.4 keV and an unabsorbed flux of
(2.4$\pm$0.3)$\times$10$^{-9}$ $\ergcms$ (2-28 keV) was estimated.  A
power law spectrum with a photon index of 1.64$\pm$0.06 gives a flux
of (2.6$\pm$0.2)$\times$10$^{-9}$ $\ergcms$ (2-28 keV; corrected for
absorption).  Converting these numbers to a mass accretion rate,
assuming standard neutron star parameters ($R$=10 km,
$M$=1.4$M_\odot$, solar abundances) and 100\% efficiency in converting
gravitational energy to radiation, gives 1.3$\times$10$^{-9}$
M$_\odot$ yr$^{-1}$. This is similar to the estimated mass accretion
rate of GX\,3$+$1 at its transition (den Hartog et~al. 2002).

\section{Discussion}

\begin{figure}[t]
\psfig{figure=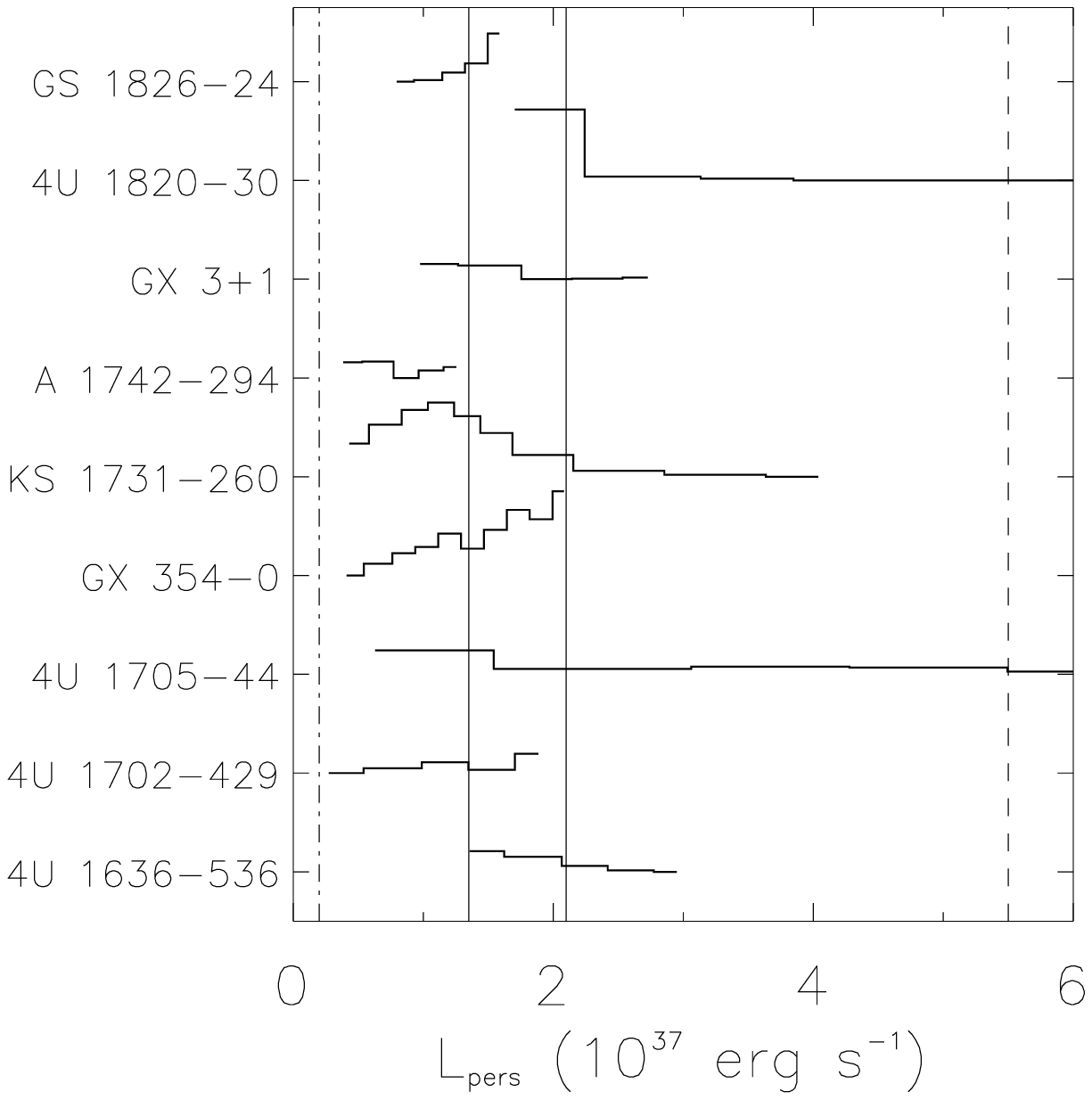,width=8cm,clip=t}
\caption{A schematic diagram of the burst rate as a function of
luminosity for the nine X-ray burst sources (see
Fig.\,\ref{burstflux}). Between the solid lines the region is indicated
where the burst rate drops by a factor of 5. No type\,I bursts were
observed at luminosities above the dashed line. The dotted-dashed line
indicates the theoretical predicted transition from hydrogen-rich
bursts to pure helium bursts.
\label{compare}}
\end{figure}

\subsection{Observational summary}

Our observations of the burst rate as a function of luminosity for the
nine frequent X-ray bursters are summarized in Fig.\,\ref{compare}. In
order to compare sources we calculated the 2--28~keV luminosity from
the flux and the distances as listed in Table\,\ref{top}. Given that
the distance has an estimated accuracy of 30\% and that the 2--28~keV
luminosity is only a crude indicator of the accretion rate, we
estimate that the luminosity correspondence between sources is
accurate to about a factor of 2. The following conclusion can be drawn
from Fig.\,\ref{compare} (and \ref{burstflux}): the burst rate shows
trends with luminosity that are consistent over all sources. Between 1
and $2\times10^{37}$~$\ergs$\ the burst rate peaks. Above that the
burst rate drops fast by roughly a factor of 5 (first observed in
GX\,3+1 by den Hartog et al. 2002, and now also seen for 4U\,1820-30,
KS\,1731-260, and 4U\,1705-44). Below that there is a smooth increase
towards the peak. Above $\simeq$5.5$\times10^{37}$~$\ergs$\ no bursts
are seen anymore. The latter is confirmed by observations of brighter
low-mass X-ray binaries in the same field with presumably similar
distances such as GX\,9+1, GX\,349+2, GX\,340+0, GX\,17+2, GX\,13+1
and GX\,5-1. None of these were seen to burst by the WFC. The general
trends in burst rate were known previously. The knowledge that our
work adds is 1) that there does seem to be a rather consistent burst
rate behavior from one burster to another, and 2) that there is a
rather discrete transition in this behavior between 1.4 to
2.1$\times10^{37}$~$\ergs$.

Our searches for (quasi-)periodicity in burst recurrence were
meaningful in three sources: GX\,354-0, KS\,1731-260 and
GS\,1826-24. The presence of quasi-periodicities is most obvious in
GS\,1826-24 (see also Ubertini et al. 1999 and Cocchi et al. 2001b),
present in KS\,1731-260, but only suggestive in GX\,354-0. The
quasi-periodicity is only present during times when the persistent
flux is below that for the peak burst rate, as is most clearly
demonstrated by KS\,1731-260 which, thanks to its transient nature,
traces a relatively wide range of fluxes. GS\,1826-24 never leaves
this domain which explains why its bursts {\em always} recur
quasi-periodically. The same appears to apply to
GX\,354-0. Quasi-periodicity has been seen previously with EXOSAT in a
number of other sources: EXO\,0748-676 (Gottwald et al. 1986),
4U\,1705-44 (Gottwald et al. 1989), Ser X-1 (Sztajno et al. 1983) and
4U\,1636-536 (Lewin et al. 1987) but only for a limited amount of
time. Our observations for the first time show empirically that the
quasi-periodicity is restricted to a very particular luminosity range
and that there is a narrow positive relationship between burst
frequency and the persistent flux.

Our determinations of burst decay times in KS\,1731-260 and
4U\,1705-44 suggest a clear correspondence between decay time, burst
rate behavior and quasi-periodicity, in the sense that there is a
clear transition at a luminosity between 1.4 and
2.1$\times10^{37}$~$\ergs$. However, the decay times observed in
GX\,3+1 and 4U\,1705-44 do not follow this trend despite tracing out
similar ranges in luminosity (formally the same applies to 4U\,1820-30
but here we know that the decay times cannot be long because the mass
donor is proven to be a helium white dwarf; Stella et~al.  1987). 

To summarize: the central finding in our study is the likely
identification of a single luminosity between 1.4 to
2.1$\times10^{37}$~$\ergs$, consistent with a single mass accretion
rate, where the bursting behavior changes in three basic ways: going
to higher luminosities, bursts 1) become rather quickly a factor of 5
less frequent, 2) stop recurring quasi-periodically, 3) stop being
long.

\subsection{Theoretical interpretation}

From theory it is expected that long bursts can only be due to helium
flashes in a hydrogen-rich environment, which is predicted to occur at
the highest or lowest accretion regimes (Fujimoto et~al. 1981;
Bildsten 1998; see also introduction).  Given the fact that in our
sample only short bursts are observed at higher luminosities, we may
identify the transition in burst behavior with the transition from the
lowest to the middle accretion regime.

At the lowest luminosities the helium flash is triggered by unstable
hydrogen burning (Fujimoto et al. 1981).  Between two bursts no
hydrogen is burned, and it is only the accretion of matter that
increases the pressure and temperature to high enough values to start
this burning. If we assume that the accretion flow is stable then the
burst wait time is only dependent on the accretion rate, and a
quasi-periodic behavior is not unexpected, as is a narrow relationship
between its frequency and the persistent flux.

In the middle accretion regime, bursts take place in a pure helium
shell which is fed by stable hydrogen burning in a layer above that.
When a critical temperature and pressure are reached the helium
ignites. Here the onset of the bursts is determined by the heating of
the shell due to the hydrogen burning and the accretion. But more
importantly, the onset of helium burning is very sensitive to the
temperature (Bildsten 1998), making it highly dependent on local
perturbations in the hydrogen burning. This means that local
conditions determine the onset (and thus the wait time) of a burst,
and quasi-periodic behavior is not readily expected anymore.

Hydrogen starts burning in an unstable fashion at lower column depths
than helium (Joss 1977). This means that the conditions for triggering
an X-ray burst in the lowest accretion regime are reached after
accreting a smaller amount of matter than in the middle accretion
regime, and a higher burst rate is expected. Assuming that only the
accretion rate can set the condition for the start of unstable
hydrogen burning, it is expected that this transition happens in a
fairly small range of accretion rates. We remark that recently
an interesting phenomenon was detected of quasi periodic
oscillations (QPOs) at 7-8 mHz in a few bursting LMXBs
which appear to coincide with this transition and whose presence is
coupled to the occurrence of type\,I bursts (Revnivtsev et
al. 2001). Revnivtsev et al. speculate that these QPOs are related to
special modes of nuclear burning. Support for this hypothesis is
provided from trends in kHz QPOs observed in one of those sources (Yu
\& van der Klis 2002).

We have for the first time shown that the largest decrease in burst
rate towards higher fluxes is coincident with the onset of stable
hydrogen burning.  Thereby, the above-mentioned explanation for the
apparently sudden decrease in burst rate may partly resolve a
long-standing problem for explaining decreasing burst rates. Van
Paradijs et~al. (1988) explained this by invoking increased stable
helium burning with increasing accretion rate. However, such burning
is not expected to occur at sub-Eddington accretion rates (Fujimoto
et~al. 1981). Bildsten (2000) suggested that non-global accretion on
the neutron star (Marshall 1982; Inogamov \& Sunyaev 1999) may explain
the decreasing burst rate: the accretion area may be smaller than the
neutron star surface so that the {\em local} mass accretion rate is
higher than the globally measured one.

It is unclear how the burst rate behaves above the transition.
Direct measurements (Fig.\,\ref{burstflux}) are ambiguous: it may be
constant or slowly decreasing. The rarity of X-ray bursts for more
luminous sources suggests that there is a decrease, but observations
in this domain are subject to strong selection effects as was already
noted by other investigators (i.e., since the persistent emission is
already close to Eddington there is hardly any room for flux increases
by bursts). If the decrease is real, that does need to be explained,
perhaps in a manner as proposed by Van Paradijs et~al. (1988) and
Bildsten (2000).

So far, we have dealt with observations that are explainable in
current burst theory. There are three observational facts for which
this is more difficult. The first is that short bursts seem
to be rather common in the low accretion regime. Long bursts would be
expected because the flashes occur in a hydrogen-rich layer. This
problem was recently also recognized by den Hartog et al. (2002) for the
specific case of GX\,3+1. Also, short type\,I bursts have been detected
with extremely low persistent fluxes (e.g., a short, 2.6 s, burst from
SAX\,J2224.9+5421 was quickly and deeply followed up in X-rays and no
persistent source was found with a 2--10~keV upper limit of
$1.3\times10^{-13}$ $\ergcms$; see Cornelisse et al. 2002). Fujimoto
et~al. (1981) briefly sketch an alternative path to trigger bursts in
the lowest regime that could explain this. If the unstable hydrogen
burning in the bottom shell does not trigger the helium burning
instantaneously it will cause temporary stable hydrogen burning in the
higher shells. The unstable hydrogen burning is not observable (Joss
1977). A pure helium layer will be built up by the stable
hydrogen burning and it will probably take a number of invisible
hydrogen flashes to trigger a helium flash in this layer, very much
like in the middle accretion regime, with a short burst as a result.

It is unclear what determines the varying mix of short and long bursts
(100\% short bursts for GX\,3+1 and GX\,354-0, less than 5\% for
GS\,1826-24, and in-between percentages for other sources). We suspect
that variability of the persistent flux may be an issue.  If we
compare the ASM lightcurves in Fig.\,\ref{asmcurves} of KS\,1731$-$260
and GS\,1826$-$24 we notice that KS\,1731$-$260 is more variable at
comparable luminosities (between MJD 51000 and MJD 51700). During the
last WFC campaign (around MJD 51800) only long bursts are observed for
KS\,1731$-$260 and the variability becomes comparable to
GS\,1826$-$24. This would indicate that low variability, (i.e., a
smooth accretion rate) gives rise to solely long X-ray bursts, while a
more inhomogeneous accretion rate gives rise to incidental pure helium
flashes, and this would lead to the following naive picture. An
inhomogeneous accretion flow would give rise to large variation
in the local accretion rate. These variations could at
certain times be high enough to start stable (or unstable) hydrogen
burning, and the hydrogen would be depleted in the burning layers.
This would give rise to hydrogen poor bursts.  For a homogeneous 
accretion flow, where the local accretion rate stays constant over
time, hydrogen burning would not start and long burst would be observed.

Another issue unexplainable by current burst theory is the
different peak burst rate between different sources. A bi-modal
distribution appears to be present.  GX\,354$-$0, KS\,1731$-$260,
4U\,1820$-$30 and GS\,1826$-$24 all have a peak burst rate of about 9
bursts day$^{-1}$, while the other sources have a peak burst rate of
about 2.5 bursts day$^{-1}$. EXOSAT observations of 4U\,1636$-$536
showed a burst rate of 8 bursts day$^{-1}$, more in line with the
first group (Lewin et~al. 1987). Given the large uncertainty in the
distance it could well be possible that during the WFC observations
the luminosity of 4U\,1636$-$536 is not low enough for such high burst
rates, and this might be the case for the other sources. We have no
good explanation for this apparent bi-modality.

As was already noted by van Paradijs (1988), there exists a large
discrepancy between theory and observations with regards to the mass
accretion rate at which the transition takes place from the lowest
accretion regime to the middle regime. We find a value of
$(1.4-2.1)\times10^{-9}$ M$_\odot$~yr$^{-1}$, thus confirming the
values found by van Paradijs et al. (1988), while theory predicts 1 to
2$\times10^{-10}$~M$_\odot$~yr$^{-1}$ (Fujimoto et al. 1981; Bildsten
1998; indicated in Fig.~5 with a dash-dotted line). The theoretical
values do depend somewhat on assumptions for a number of neutron star
parameters (radius, mass, etc.) and conditions in the neutron star
(abundances of CNO elements, opacity, etc.) and become somewhat
uncertain due to uncertainties in the distance. This results in
substantial margins in the theoretical value from source to source of
a factor of three. Still, this is insufficient to explain the
discrepancy.

Bildsten (2000) suggests that the discrepancy can be resolved if the
transition is not from the {\em lower} to the middle regime but from
the {\em upper} to the middle regime. This may be accomplished if the
area over which the accretion takes place is smaller than the neutron
star and grows faster than linearly with the global accretion
rate. Thus, the mass accretion rate per unit area diminishes as the
global accretion rates grows and the source moves from the middle to
the lower regime with increasing persistent fluxes. This idea is based
on a suggestive radius versus persistent flux trend observed in
EXO~0748-676 (Gottwald et al. 1986). It remains to be seen from a
complete analysis of our sample, involving time-resolved spectroscopy
of all bursts, whether the radius shows such trends with persistent
flux. However, a partial result by den Hartog et al. (2003) on GX~3+1
shows no indications for such a trend. Furthermore, the measurements
of one source in particular strongly suggest that this effect is not
important: KS 1731-260 is the only frequent bursters that exhibits a
large dynamic range in persistent flux due to it being transient. The
source gradually dies out in our observations and traverses from high
global mass accretion rates to a negligible ones. Our observations show
that once it exited the middle regime, it never changed regime
again. It seems unlikely that KS~1731-260 remains in the upper mass
accretion rates {\em per unit area} regime all the way down to
negligible {\em global} accretion rates. Therefore, our data suggest
that the explanation as put forward by Bildsten (2000) for the
discrepancy in predicted and observed threshold value between lower
and middle accretion burst regime, cannot be general. This is further
supported by the quasi-periodicity that we find in the burst
recurrence for KS~1731-260 in this regime as well as GS~1826-24. Such
a form of stability is more likely to happen if the trigger criterion
for the flash is dependent on fewer parameters. In the lower burst
regime, where the hydrogen is first ignited, the trigger is only
dependent on pressure while in the upper regime, where the helium
first ignites, it also depends on the local temperature.

\section{Summary}

Thanks to the wide (50\%) and long (7~Ms) coverage of the population
of low-mass X-ray binaries we detected an unprecedented large number
of type\,I X-ray bursts for nine sources which enabled us to perform a
comparative study of bursting behavior which was hitherto not
possible. We were able to accurately detect systematic trends in burst
rate, burst duration and burst recurrence periodicity and to put them
under the common denominator provided by current burst theory. Our
central finding is that most of the trends in bursting behavior are
driven by the onset of stable hydrogen burning in the neutron star
atmosphere. Furthermore, we notice three new observational facts which
are more difficult to explain with current burst theory: the presence
of short pure-helium bursts at the lowest accretion regimes, the
bimodal distribution of peak burst rates, and an accretion rate at
which the onset of stable hydrogen burning occurs that is ten times
higher than predicted. Finally, we note that our investigation is the
first to signal quasi-periodic burst recurrence in KS\,1731-260, and a
clear proportionality between the frequency of the
quasi-periodicity and the persistent flux in GS\,1826-24 and
KS\,1731-260.

\end{document}